\def\tr{{\rm tr}\,}
\def\Tr{{\rm Tr}\,}
\def\wt{\widetilde}
\def\sgn{{\rm sgn\,}}
\def\b{\bibitem}
\def\be{\begin{equation}}
\def\ee{\end{equation}}
\def\bea{\begin{eqnarray}}
\def\eea{\end{eqnarray}}
\def\bml{\begin{mathletters}}
\def\eml{\end{mathletters}}
\documentstyle[aps,prb,eqsecnum,psfig,epsf,floats]{revtex}
\draft
\begin{document}
% Macros for the various macro package names, etc.
\def\SNG{{\em Physical Review Style and Notation Guide}}
\def\LUG {{\em \LaTeX{} User's Guide \& Reference Manual}}
\def\btt#1{{\tt$\backslash$\string#1}}%
\def\REVTeX{REV\TeX}
\def\AmS{{\protect\the\textfont2
        A\kern-.1667em\lower.5ex\hbox{M}\kern-.125emS}}
\def\AmSLaTeX{\AmS-\LaTeX}
\def\BibTeX{\rm B{\sc ib}\TeX}
%\makeatletter
%\tighten
\twocolumn[\hsize\textwidth\columnwidth\hsize\csname@twocolumnfalse%
\endcsname

\title{Theory of Disordered Itinerant Ferromagnets I: Metallic Phase
%      \small{$[$ Phys. Rev. B {\bf xx}, xxxx (1999) $]$}
}
\author{T.R.Kirkpatrick}
\address{Institute for Physical Science and Technology, and Department of
 Physics\\
 University of Maryland,\\
 College Park, MD 20742}
\author{D.Belitz}
\address{Department of Physics and Materials Science Institute\\
University of Oregon,\\
Eugene, OR 97403}
\date{\today}
\maketitle

\begin{abstract}
A comprehensive theory for electronic transport in itinerant ferromagnets
is developed. We first show that the $Q$-field theory used previously to
describe a disordered Fermi liquid also has a saddle-point solution that
describes a ferromagnet in a disordered Stoner approximation. We
calculate transport coefficients and thermodynamic susceptibilities by
expanding about the saddle point to Gaussian order. At this level, the
theory generalizes previous RPA-type theories by including quenched 
disorder. We then study soft-mode effects in the ferromagnetic 
state in a one-loop approximation. In three-dimensions, we find that the 
spin waves induce a square-root frequency dependence of the conductivity,
but not of the density of states, that is qualitatively the same as the
usual weak-localization effect induced by the diffusive soft modes. In
contrast to the weak-localization anomaly, this effect persists also at
nonzero temperatures.
In two-dimensions, however, the spin waves do not lead to a logarithmic
frequency dependence. This explains experimental observations in thin 
ferromagnetic films, and it provides a basis for the construction of a 
simple effective field theory for the transition from a ferromagnetic
metal to a ferromagnetic insulator.
\end{abstract}
]
%\narrowtext
%\tableofcontents

\section{Introduction}
\label{sec:I}

The theoretical treatment of many-fermions systems is a hard problem,
especially in the presence of quenched disorder. Traditional approaches
have been Landau Fermi-liquid theory,\cite{BaymPethick}
and many-body perturbation theory.\cite{FetterWalecka,AGD}
Impressive progress has been made within the framework of the latter.
The random phase approximation (RPA) has been developed for the
Fermi liquid phase, and similar theories have been used to describe
magnetic and superconducting phases. In a more recent development,
the corrections to Fermi-liquid theory known as 
``weak-localization effects''\cite{weak_localization_footnote}
have been derived using many-body diagrammatic techniques.\cite{LeeRama}
Even more recently, field-theoretic methods have been applied 
to the many-fermion problem, which have certain advantages over the 
traditional techniques. Most importantly, they allow for a straightforward 
application of the renormalization group (RG), implementing,
{\em inter alia}, an old program 
of describing the various phases of many-body systems in terms of stable RG 
fixed points.\cite{Shankar} So far this program has been carried out for 
clean and disordered Fermi liquids.\cite{Shankar_ff}
It has also been shown that the weak-localization effects 
can be understood as the leading corrections to scaling near a disordered 
Fermi-liquid fixed point.\cite{us_fermions} A further advantage of the
field theoretic approach is that it allows for a systematic identification
and analysis of the soft or massless modes that are
responsible for the long-distance, long-time properties of the system.
If desirable, it also allows for the derivation of effective theories that
keep only the soft modes explicitly, while integrating out all other
degrees of freedom in some simple approximation. This ability to focus
on soft modes is
particularly important for studies of the zero-temperature ($T=0$)
properties of many-body systems, since it turns out that there are
many soft modes at $T=0$ that acquire a mass at nonzero temperature.
Conservation laws, and analogies to classical fluids, are therefore of
limited value in identifying the soft modes of a quantum system at $T=0$.

An example of soft modes that have no analogs in classical physics are
the diffusive modes that cause the weak-localization effects mentioned above, 
which are long-wavelength/low-frequency nonanalyticities in the frequency and
wavenumber dependence of transport coefficients and thermodynamic
quantities. These nonanalyticities, which generically take the form of
power laws, are examples of a more general phenomenon known as generic
scale invariance. Correlation functions which, as functions of space and time,
are power laws and thus homogeneous functions (as opposed to, e.g., 
exponentials), contain no intrinsic length or time scales, hence the name
``scale invariance''. The most widely known example of scale invariance,
which is caused by soft modes, occurs at critical points as a result of the
critical modes. Critical points are exceptional points in a
phase diagram, and reaching them requires a fine tuning of parameters.
{\em Generic} scale invariance, on the other hand, does not require any
fine tuning and is due to soft modes that occur over large
regions of parameter space, like the diffusive ``diffuson'' modes in disordered
electron sytems that cause the weak-localization effects. However, generic 
scale invariance is in no way restricted to quantum systems; it has been
discussed primarily in classical systems ranging from classical fluids and
liquid crystals to sandpiles.\cite{GSI}

It is the purpose of the present paper to develop a comprehensive field 
theoretic method for describing disordered itinerant ferromagnets, 
which so far have not been studied with these techniques, with particular
emphasis on the consequences of the soft modes in such systems.
Specifically, we generalize the theory of Ref.\ \onlinecite{us_fermions}
to the magnetic case. Simple approximations within this theory yield
results that correspond to disordered Stoner theory for the equation of
state, and to disordered RPA-type approximations for the transport
properties. We then study generic scale invariance effects on the background
of ferromagnetism. An interesting question in this context
is the role of the Goldstone modes or spin waves that arise from the 
spontaneously broken spin rotational symmetry, which represent additional
soft modes compared to a system without magnetic long-range order.
Since the soft mode structure of ferromagnets
is thus different from that of nonmagnetic systems, one also expects
generic scale invariance phenomena that differ from the usual
weak-localization nonanalyticities. Indeed, we find that the spin waves
contribute to the leading nonanalyticitic frequency dependence of the
conductivity in three-dimensions, but not to that of the density of
states. Also, we find that they do not lead to a logarithmic frequency
dependence of the conductivity in two-dimensions. This
explains experimental observations that have found the frequency anomaly
in thin ferromagnetic films to be the same as those in nonmagnetic 
metals in an external magnetic field. 

This paper is organized as follows. In Sec.\ \ref{sec:II} we recall the
general $Q$-matrix field theory for disordered interacting electrons that
was developed in Ref.\ \onlinecite{us_fermions}, and we show that this 
theory allows for a saddle-point solution that corresponds to a disordered 
Stoner theory. In Sec.\ \ref{sec:III} we calculate the spin and density
susceptibilities, as well as the conductivity, in a Gaussian approximation.
This produces generalizations of well-known results to the case of
disordered magnets. In Sec.\ \ref{sec:IV} we proceed
to perform a one-loop calculation
of the density of states and the conductivity, and calculate the contribution
of the Goldstone modes to the leading nonanalyticities.
In Sec.\ \ref{sec:V} we conclude with a discussion of our results.
The ferromagnetic-metal-to-ferromagnetic-insulator transition will
be discussed in a second paper,\cite{us_paper_II} which we will refer to
as (II).

\section{Matrix field theory}
\label{sec:II}

\subsection{$Q$-matrix theory for fermions}
\label{subsec:II.A}

We start with a general field theory for electrons. For any
fermionic system, the partition function can be written as a functional
integral over fermionic (i.e., Grassmann valued)
fields $\bar\psi$ and $\psi$,\cite{NegeleOrland}
\begin{mathletters}
\label{eqs:2.1}
\begin{equation}
Z = \int D[{\bar\psi},\psi]\ \exp\left(S[{\bar\psi},
                       \psi] \right)\quad,
\label{eq:2.1a}
\end{equation}
where $S$ is the action. We consider an action that consists of a
free-fermion part $S_0$, a part $S_{\rm dis}$ describing the interaction
of the electrons with quenched disorder, and a part $S_{\rm int}$ describing
the electron-electron interaction,
\begin{equation}
S = S_0 + S_{\rm dis} + S_{\rm int}\quad,
\label{eq:2.1b}
\end{equation}
\end{mathletters}%
Each field $\psi$ or $\bar\psi$ carries a Matsubara frequency
index $n$ and a spin index $\sigma=\uparrow,\downarrow\equiv +,-$. 
Since we will deal
with the quenched disorder by means of the replica trick, each field also
carries a replica index $\alpha$. It is useful to 
introduce a matrix of bilinear products of the fermion fields,
\begin{eqnarray}
B_{12} &=& \frac{i}{2}\,\left( \begin{array}{cccc}
          -\psi_{1\uparrow}{\bar\psi}_{2\uparrow} &
             -\psi_{1\uparrow}{\bar\psi}_{2\downarrow} &
                 -\psi_{1\uparrow}\psi_{2\downarrow} &
                      \ \ \psi_{1\uparrow}\psi_{2\uparrow}  \\
          -\psi_{1\downarrow}{\bar\psi}_{2\uparrow} &
             -\psi_{1\downarrow}{\bar\psi}_{2\downarrow} &
                 -\psi_{1\downarrow}\psi_{2\downarrow} &
                      \ \ \psi_{1\downarrow}\psi_{2\uparrow}  \\
          \ \ {\bar\psi}_{1\downarrow}{\bar\psi}_{2\uparrow} &
             \ \ {\bar\psi}_{1\downarrow}{\bar\psi}_{2\downarrow} &
                 \ \ {\bar\psi}_{1\downarrow}\psi_{2\downarrow} &
                      -{\bar\psi}_{1\downarrow}\psi_{2\uparrow} \\
          -{\bar\psi}_{1\uparrow}{\bar\psi}_{2\uparrow} &
             -{\bar\psi}_{1\uparrow}{\bar\psi}_{2\downarrow} &
                 -{\bar\psi}_{1\uparrow}\psi_{2\downarrow} &
                      \ \ {\bar\psi}_{1\uparrow}\psi_{2\uparrow} \\
                    \end{array}\right)
\nonumber\\
&\cong& Q_{12}\quad.
\label{eq:2.2}
\end{eqnarray}
where all fields are understood to be taken at position ${\bf x}$, and
$1\equiv (n_1,\alpha_1)$, etc. The
matrix elements of $B$ commute with one another, and are therefore
isomorphic to classical or number-valued fields that we denote by 
$Q$.\cite{Notation_Footnote}
This isomorphism maps the adjoint operation on products of fermion fields,
which is denoted above by an overbar, on the complex conjugation of the
classical fields. We use the isomorphism to
constrain $B$ to the classical field $Q$, and exactly rewrite the partition
function\cite{us_fermions}
\begin{eqnarray}
Z &=& \int D[{\bar\psi},\psi]\ e^{S[{\bar\psi},\psi]}
      \int D[Q]\,\delta[Q-B]
\nonumber\\
  &=& \int D[{\bar\psi},\psi]\ e^{S[{\bar\psi},\psi]}
      \int D[Q]\,D[{\wt\Lambda}]\ e^{\Tr [{\wt\Lambda}(Q-B)]}
\nonumber\\
  &\equiv& \int D[Q]\,D[{\wt\Lambda}]\ e^{{\cal A}[Q,{\wt\Lambda}]}\quad.
\label{eq:2.3}
\end{eqnarray}
We have introduced an
auxiliary bosonic matrix field ${\wt\Lambda}$ to enforce the
functional delta-constraint in the first line of Eq.\ (\ref{eq:2.3}),
and the last line defines the action ${\cal A}$.
The matrix elements of both $Q$ and ${\wt\Lambda}$
are spin-quaternions, i.e. elements of ${\cal Q}\times{\cal Q}$ with
${\cal Q}$ the quaternion field. From Eq.\ (\ref{eq:2.2}) we see that
expectation values of the $Q$ matrix elements yield 
single-particle Green functions, and $Q$-$Q$
correlation functions describe four-fermion correlation functions.
The physical meaning of ${\wt\Lambda}$ 
is that its expectation value plays the role of a
self energy (see Ref.\ \onlinecite{us_fermions} and Sec.\ \ref{subsec:II.B}
below). 

It is convenient to expand the $4\times 4$ matrix in Eq.\ (\ref{eq:2.2})
in a spin-quaternion basis,
\begin{equation}
Q_{12}({\bf x}) = \sum_{r,i=0}^{3} (\tau_r\otimes s_i)\,{^i_rQ_{12}}({\bf x})
                 \quad
\label{eq:2.4}
\end{equation}
and analogously for $\wt\Lambda$. Here 
$\tau_0 = s_0 = \openone_2$ is the
$2\times 2$ unit matrix, and $\tau_j = -s_j = -i\sigma_j$, $(j=1,2,3)$,
with $\sigma_{1,2,3}$ the Pauli matrices. In this basis, $i=0$ and $i=1,2,3$
describe the spin singlet and the spin triplet, respectively. An explicit
calculation using Eq.\ (\ref{eq:2.2}) reveals that $r=0,3$ corresponds 
to the particle-hole channel
(i.e., products ${\bar\psi}\psi$), while $r=1,2$ describes the
particle-particle channel (i.e., products ${\bar\psi}{\bar\psi}$ or
$\psi\psi$). 
From the structure of Eq.\ (\ref{eq:2.2}) one obtains the
following formal symmetry properties of the $Q$ matrices,\cite{us_fermions}
\begin{mathletters}
\label{eqs:2.5}
\begin{eqnarray}
{^0_r Q}_{12}&=&(-)^r\,{^0_r Q}_{21}\quad\quad (r=0,3)\quad,
\label{eq:2.5a}\\
{^i_r Q}_{12}&=&(-)^{r+1}\,{^i_r Q}_{21}\quad (r=0,3;\ i=1,2,3)\quad,
\label{eq:2.5b}\\
{^0_r Q}_{12}&=&{^0_r Q}_{21}\quad\quad\qquad\, (r=1,2)\quad,
\label{eq:2.5c}\\
{^i_r Q}_{12}&=&-{^i_r Q}_{21}\quad\quad\quad\ (r=1,2;\ i=1,2,3)\quad,
\label{eq:2.5d}\\
{^i_r Q}_{12}^*&=&- {^i_r Q}_{-n_1-1,-n_2-1}^{\alpha_1\alpha_2}\quad.
\label{eq:2.5e}
\end{eqnarray}
\end{mathletters}%
The star in Eq.\ (\ref{eq:2.5e}) denotes complex conjugation.

For the purposes of the present paper, we will be particularly interested
in the matrix elements $^0_0 Q$ and $^3_3 Q$. From Eqs.\ (\ref{eq:2.2}) and
(\ref{eq:2.4}) we find
\bml
\bea
\langle\, {^0_0 Q}_{12}({\bf x})\rangle &\cong& \delta_{12}\ 
  \frac{i}{4}\sum_{\sigma} \langle{\bar\psi}_{1\sigma}({\bf x})\,
     \psi_{1\sigma}({\bf x})\rangle\quad,
\label{eq:2.6a}\\
\langle\, {^3_3 Q}_{12}({\bf x})\rangle &\cong& \delta_{12}\ 
  \frac{i}{4}\sum_{\sigma}\sigma\ \langle{\bar\psi}_{1\sigma}({\bf x})\,
     \psi_{1\sigma}({\bf x})\rangle\quad,
\label{eq:2.6b}
\eea
\eml%
where $\langle\ldots\rangle$ denotes an average taken with the full action.
From these expressions we obtain various observables in terms of expectation
values of the $Q$ fields, for instance the particle number density $n$, 
the frequency or
energy dependent density of states $N$, with energy measured from the
chemical potential $\mu$ (or, at $T=0$, from the Fermi energy 
$\epsilon_{\rm F}$), and the magnetization $M$,
\bml
\label{eqs:2.7}
\bea
n &=& -4i\,T\sum_n 
      \bigl\langle\, {^0_0 Q}_{nn}^{\alpha\alpha}({\bf x})\bigr\rangle\quad,
\label{eq:2.7a}\\
N(\mu + \omega) &=& \frac{4}{\pi}\ {\rm Re}\ 
    \langle\, {^0_0 Q}_{nn}^{\alpha\alpha}({\bf x})\rangle
        {\bigl\vert}_{i\omega_n\rightarrow\omega+i0}\quad,
\label{eq:2.7b}\\
M &=& -4i\mu_{\rm B}\,T\sum_n
      \bigl\langle\, {^3_3 Q}_{nn}^{\alpha\alpha}({\bf x})\bigr\rangle\quad,
\label{eq:2.7c}
\eea
\eml%
with $\mu_{\rm B}$ the Bohr magneton. Here and in what follows we denote
fermionic Matsubara frequencies by $\omega_n = 2\pi T(n+1/2)$, and bosonic
ones by $\Omega_n = 2\pi Tn$. We use units such that $\hbar = k_{\rm B} = 1$.

By using the delta constraint in Eq.\ (\ref{eq:2.3}) to rewrite all terms 
that are quartic in the fermion field in terms of $Q$, we can achieve
an integrand that is bilinear in $\psi$ and $\bar\psi$. The Grassmannian
integral can then be performed exactly, and we obtain for the
action ${\cal A}$
\begin{mathletters}
\label{eqs:2.8}
\begin{eqnarray}
{\cal A}[Q,{\wt\Lambda}] &=& {\cal A}_{\rm dis} + {\cal A}_{\rm int}
                           + \frac{1}{2}\,\Tr\ln\left(G_0^{-1} - i{\wt\Lambda}
                                       \right)
\nonumber\\
  && + \int d{\bf x}\ \tr\left({\wt\Lambda}({\bf x})\,Q({\bf x})\right)\quad.
\label{eq:2.8a}
\end{eqnarray}
Here
\begin{equation}
G_0^{-1} = -\partial_{\tau} + \partial_{\bf x}^2/2m_{\rm e} + \mu\quad,
\label{eq:2.8b}
\end{equation}
\end{mathletters}%
is the inverse free electron Green operator, with $\partial_{\tau}$ and
$\partial_{\bf x}$ derivatives with respect to imaginary time and position,
respectively, and $m_{\rm e}$ is the electron mass.
$\Tr$ denotes a trace over all degrees of freedom, including the continuous
position variable, while $\tr$ is a trace over all those discrete indices that
are not explicitly shown. 
For the disorder part of the action one finds\cite{us_fermions}
\be
{\cal A}_{\rm dis}[Q] = \frac{1}{\pi N_F\tau}\int d{\bf x}\
                       \tr \bigl(Q({\bf x})\bigr)^2\quad.
\label{eq:2.9}
\ee
with $N_{\rm F}$ the density of states at the Fermi level in saddle-point
approximation (see Ref.\ \onlinecite{us_fermions} and Sec.\ \ref{subsec:III.C}
below), and
$\tau$ the single-particle scattering or relaxation time.\cite{dis_footnote}
The electron-electron interaction ${\cal A}_{\rm int}$ is conveniently 
decomposed into four pieces that describe the interaction
in the particle-hole and particle-particle spin-singlet and spin-triplet 
channels.\cite{us_fermions} For reasons that will be explained in
Sec.\ \ref{subsec:III.B} below, for our purposes we can neglect the
particle-particle channel. We thus drop $r=1,2$ from our spin-quaternion
basis, Eq.\ (\ref{eq:2.4}). In particular, we write the interaction as
the sum of the particle-hole spin-singlet and triplet terms,
\begin{mathletters}
\label{eqs:2.10}
\begin{equation}
{\cal A}_{\rm int}[Q] = {\cal A}_{\rm int}^{\,(s)} 
     + {\cal A}_{\rm int}^{\,(t)}\quad,
\label{eq:2.10a}
\end{equation}
\begin{eqnarray}
{\cal A}_{\rm int}^{\,(s)}&=&\frac{T\Gamma^{(s)}}{2}\int d{\bf x}
              \sum_{r=0,3}(-1)^r \sum_{n_1,n_2,m}\sum_\alpha
\nonumber\\
&&\times\left[\tr \left((\tau_r\otimes s_0)\,Q_{n_1,n_1+m}^{\alpha\alpha}
({\bf x})\right)\right]
\nonumber\\
&&\times\left[\tr \left((\tau_r\otimes s_0)\,Q_{n_2+m,n_2}^{\alpha\alpha}
({\bf x})\right)\right]\quad,
\label{eq:2.10b}
\end{eqnarray}
\begin{eqnarray}
{\cal A}_{\rm int}^{\,(t)}&=&\frac{T\Gamma^{(t)}}{2}\int d{\bf x}
    \sum_{r=0,3}(-1)^r \sum_{n_1,n_2,m}\sum_\alpha\sum_{i=1}^3
\nonumber\\
&&\times\left[\tr\left((\tau_r\otimes s_i)\,Q_{n_1,n_1+m}^{\alpha\alpha}
({\bf x})\right)\right]
\nonumber\\
&&\times\left[\tr\left((\tau_r\otimes s_i)\,Q_{n_2+m,n_2}^{\alpha\alpha}
({\bf x})\right)\right]\quad.
\label{eq:2.10c}
\end{eqnarray}
\eml%
Here $\Gamma^{(s)}>0$ and $\Gamma^{(t)}>0$ are the spin-singlet and 
spin-triplet interaction amplitudes, respectively. $\Gamma^{(t)}$ is
responsible for producing magnetism.

\subsection{The Stoner saddle point}
\label{subsec:II.B}

We now look for a saddle-point solution of the field theory derived in
the previous subsection. We are interested in an itinerant ferromagnet,
i.e. a state where both the density of states at the Fermi level and
the magnetization are nonzero. We are further looking for a homogeneous
state, so we drop the real space dependence of the fields.
Equations (\ref{eqs:2.7}) suggest the
following {\it ansatz} for the saddle-point fields,
\bml
\label{eqs:2.11}
\bea
{^i_r Q}_{12}{\bigl\vert}_{\rm sp}&=&\delta_{12}\,\left[\delta_{r0}\,
     \delta_{i0}\,G_{n_1} + \delta_{r3}\,\delta_{i3}\,F_{n_1}
          \right]\quad,
\label{eq:2.11a}\\
{^i_r {\wt \Lambda}}_{12}{\bigl\vert}_{\rm sp}&=&\delta_{12}\,
       \left[-\delta_{r0}\,
     \delta_{i0}\,i\Sigma_{n_1} + \delta_{r3}\,\delta_{i3}\,
          i\Delta_{n_1}\right]\quad.
\label{eq:2.11b}
\eea
\eml%
Substituting this into Eqs.\ (\ref{eqs:2.8}) - (\ref{eqs:2.10}), and
using the saddle-point condition $\delta {\cal A}/\delta Q =
\delta {\cal A}/\delta {\wt\Lambda} = 0$, we obtain the saddle-point
equations
\bml
\label{eqs:2.12}
\bea
G_n&=&\frac{i}{2V}\sum_{\bf k} {\cal G}_n({\bf k})\quad,
\label{eq:2.12a}\\
F_n&=&\frac{i}{2V}\sum_{\bf k} {\cal F}_n({\bf k})\quad,
\label{eq:2.12b}\\
\Sigma_n&=&\frac{-2i}{\pi N_{\rm F}\tau}\ G_n - 4i\,\Gamma^{(s)}\,T
                \sum_m e^{i\omega_m 0}\,G_m\quad,
\label{eq:2.12c}\\
\Delta_n&=&\frac{2i}{\pi N_{\rm F}\tau}\ F_n - 4i\,\Gamma^{(t)}\,T
                \sum_m e^{i\omega_m 0}\,F_m\quad,
\label{eq:2.12d}
\eea
\eml%
Here
\bml
\label{eqs:2.13}
\bea
{\cal G}_n({\bf k})&=&\frac{1}{2}\,\left[{\cal G}_n^+({\bf k}) 
     + {\cal G}_n^-({\bf k})\right] \quad,
\label{eq:2.13a}\\
{\cal F}_n({\bf k})&=&\frac{1}{2}\,\left[{\cal G}_n^+({\bf k}) 
     - {\cal G}_n^-({\bf k})\right] \quad,
\label{eq:2.13b}
\eea
are Green functions given in terms of
\be
{\cal G}_n^{\pm}({\bf k}) = \frac{1}{i\omega_n - \xi_{\bf k} \pm \Delta_n
                               - \Sigma_n}\quad,
\label{eq:2.13c}
\ee
with $\xi_{\bf k} = {\bf k}^2/2m - \mu$. For later reference we also define
a matrix saddle-point Green function 
\be
G_{\rm sp} = \left(G_0^{-1} - i{\wt\Lambda}\right)^{-1}\biggr\vert_{\rm sp}
                \quad,
\label{eq:2.13d}
\ee
whose matrix elements are given by
\be
(G_{\rm sp})_{12} = \delta_{12}\ \left[{\cal G}_{n_1}\,(\tau_0\otimes s_0)
   + {\cal F}_{n_1}\,(\tau_3\otimes s_3)\right]\quad.
\label{eq:2.13e}
\ee
\eml%

Clearly, $\Sigma$ and $\Delta$ are self energies that describe the interaction
in Hartree-Fock approximation, and the disorder in self-consistent Born
approximation. The Green functions ${\cal G}$ and ${\cal F}$ obey the equations
\bml
\label{eqs:2.14}
\bea
\left(i\omega_n - \xi_{\bf k} - \Sigma_n\right)\ {\cal G}_n({\bf k})
   + \Delta_n\,{\cal F}_n({\bf k})&=&1\quad,
\label{eq:2.14a}\\
\left(i\omega_n - \xi_{\bf k} - \Sigma_n\right)\ {\cal F}_n({\bf k})
   + \Delta_n\,{\cal G}_n({\bf k})&=&0\quad.
\label{eq:2.14b}
\eea
\eml%
If we absorb the second, Hartree-Fock, contribution to $\Sigma$ in
Eq.\ (\ref{eq:2.12c}), which
is purely real and frequency independent, into a redefinition of the 
chemical potential, we can
rewrite Eqs.\ (\ref{eq:2.12c},\ref{eq:2.12d}) in the form
\bea
\Sigma_n&=&\frac{1}{\pi N_{\rm F}\tau}\ 
           \frac{1}{V}\sum_{\bf k}{\cal G}_n({\bf k})\quad,
\eqnum{2.12c'}\\
\Delta_n&=&\Delta - \frac{1}{\pi N_{\rm F}\tau}\,\frac{1}{V}\sum_{\bf k}
                    {\cal F}_n({\bf k})\quad,
\eqnum{2.12d'}
\eea
where
\be
\Delta = 2\Gamma^{(t)}T\sum_n \frac{1}{V}\sum_{\bf k}{\cal F}_n({\bf k})\quad.
\label{eq:2.15}
\ee
From Eqs.\ (\ref{eq:2.7c}), (\ref{eqs:2.11}), (\ref{eqs:2.12}), and 
(\ref{eq:2.15}) we see that $\Delta$ is
related to the magnetization in saddle-point approximation, 
$\Delta = \Gamma^{(t)}\,M/\mu_{\rm B}$.

To discuss our saddle-point solution, let us first consider the clean limit,
$\tau=\infty$, $\Sigma_n = 0$, $\Delta_n = \Delta$. From Eqs.\ (\ref{eqs:2.14},
\ref{eq:2.15}) we then obtain an expression for the equation of state that is
familiar from Stoner theory,\cite{Moriya}
\be
1 = -2\Gamma^{(t)}\,T\sum_n \frac{1}{V}\sum_{\bf k}
     \frac{1}{(i\omega_n - \xi_{\bf k})^2 - \Delta^2}\quad.
\label{eq:2.16}
\ee
In particular, the threshold value of $\Gamma^{(t)}$ for the onset of
ferromagnetism is given by the Stoner criterion
\be
N_{\rm F}\,\Gamma^{(t)} = 1\quad,
\label{eq:2.17}
\ee
and the usual physical interpretation of Stoner theory applies.\cite{Moriya}
For instance, the ${\cal G}_n^{\pm}({\bf k})$ are Hartree-Fock Green functions
with the chemical potential shifted by $\pm\Delta_n$, which is essentially
the magnetization.

The discussion of the disordered case proceeds in exact analogy to the
BCS-Gorkov theory of disordered superconductors (this is the main reason
why we have chosen to write Stoner theory in a form analogous to BCS
theory).\cite{AGD} The equation of state then takes the form
\be
1 = -2\Gamma^{(t)}\,T\sum_n \frac{1}{V}\sum_{\bf k}
     \frac{1}{\left(i\omega_n-\xi_{\bf k}+\frac{i}{2\tau}\sgn\omega_n\right)^2 
                  - \Delta^2}\quad.
\label{eq:2.18}
\ee
The integral is independent of the disorder to lowest order in $1/\tau$,
and so is therefore the Stoner criterion, Eq.\ (\ref{eq:2.17}). This is
the magnetic analog of Anderson's theorem in superconductivity.

We conclude that our saddle-point solution of the field theory, Eqs.\ (\ref{eqs:2.8}) - (\ref{eqs:2.10}), describes a ferromagnetic state in a disordered Stoner
approximation that is analogous to BCS-Gorkov theory for disordered
superconductors.

\section{Gaussian approximation}
\label{sec:III}

\subsection{Gaussian action}
\label{subsec:III.A}

We now write $Q = Q_{\rm sp} + \delta Q$, and 
${\wt\Lambda} = {\wt\Lambda}_{\rm sp} + \delta{\wt\Lambda}$, and expand
the action, Eq.\ (\ref{eq:2.8a}), in powers of $\delta Q$ and
$\delta{\wt\Lambda}$. To Gaussian order we obtain ${\cal A} = {\cal A}_{\rm sp}
+ {\cal A}_{\rm G}$ with
\bea
{\cal A}_{\rm G}&=&{\cal A}_{\rm dis}[\delta Q] + {\cal A}_{\rm int}[\delta Q]
   + \frac{1}{4}\,\Tr\left(G_{\rm sp}\,\delta{\wt\Lambda}\,G_{\rm sp}\,
     \delta{\wt\Lambda}\right)
\nonumber\\
&& + \Tr(\delta{\wt\Lambda}\,\delta Q)\quad.
\label{eq:3.1}
\eea
Defining Fourier transforms of the fields, ${\wt\Lambda}({\bf k}) =
\int d{\bf x}\,\exp(i{\bf k}{\bf x})\,{\wt\Lambda}({\bf x})$, and
analogously for $Q$, the
piece of ${\cal A}_{\rm G}$ that is quadratic in $\wt\Lambda$ can be
written
\bml
\label{eqs:3.2}
\be
\frac{1}{V}\sum_{\bf k}\sum_{1234}\sum_{rs}\sum_{ij}
   {^i_r(}\delta{\wt\Lambda)}_{12}
   ({\bf k})\ {^{ij}_{rs}A}_{12,34}({\bf k})\ 
         {^j_s(}\delta{\wt\Lambda)}_{34}(-{\bf k})\ ,
\label{eq:3.2a}
\ee
with the matrix $A$ given by
\bea
{^{ij}_{rs}A}_{12,34}({\bf k})&=&\delta_{13}\,\delta_{24}\,\left[\right.
   \varphi^{00}_{n_1n_2}({\bf k})\,m^{00}_{rs,ij} +
   \varphi^{03}_{n_1n_2}({\bf k})\,m^{03}_{rs,ij}
\nonumber\\
&&\hskip 10pt + \varphi^{30}_{n_1n_2}({\bf k})\,m^{30}_{rs,ij} +
   \varphi^{33}_{n_1n_2}({\bf k})\,m^{33}_{rs,ij}\left.\right]
\nonumber\\
&\equiv& \delta_{13}\,\delta_{24}\,{^{ij}_{rs}A}_{12}({\bf k})\quad.
\label{eq:3.2b}
\eea
\eml%
Here we have defined convolutions of Green functions
\bml
\label{eqs:3.3}
\be
\varphi^{ij}_{nm}({\bf k}) = \frac{1}{V}\sum_{\bf p} G^i_n({\bf p})\,
   G^j_m({\bf p}+{\bf k})\quad,
\label{eq:3.3a}
\ee
where $i,j=0,3$ and $G^0\equiv{\cal G}$, $G^3\equiv{\cal F}$. The matrices
$m^{00}$ etc. are defined as
\bea
m^{00}_{rs,ij} &=& \frac{1}{4}\,\tr (\tau_r\tau_s^{\dagger})\,
                              \tr (s_i s_j^{\dagger})\quad,
\label{eq:3.3b}\\
m^{03}_{rs,ij} &=& \frac{1}{4}\,\tr (\tau_r\tau_3\tau_s^{\dagger})\,
                 \tr (s_3 s_i s_j^{\dagger})\quad,
\label{eq:3.3c}\\
m^{30}_{rs,ij} &=& \frac{1}{4}\,\tr (\tau_r\tau_3\tau_s^{\dagger})\,
                 \tr (s_i s_3 s_j^{\dagger})\quad,
\label{eq:3.3d}\\
m^{33}_{rs,ij} &=& \frac{1}{4}\,\tr (\tau_r\tau_s^{\dagger})\,
                 \left({{+\atop +}\atop{+\atop -}}\right)_i\,
                 \tr (s_i s_j^{\dagger})\quad,
\label{eq:3.3e}
\eea
\eml%
where $\left({{+\atop +}\atop{+\atop -}}\right)_i = \delta_{i0}
+ \delta_{i1} + \delta_{i2} - \delta_{i3}$, etc.
$Q$ and $\wt\Lambda$ can now be decoupled by shifting and scaling the
$\wt\Lambda$ field. If we define a new field ${\bar\Lambda}$ by
\be
\delta{\wt\Lambda}({\bf k}) = 2A^{-1}\left(\delta{\bar\Lambda}({\bf k}) 
   - \delta Q({\bf k})\right)\quad,
\label{eq:3.4}
\ee
then $\bar\Lambda$ and $Q$ decouple. We can thus integrate out 
$\delta{\bar\Lambda}$ and obtain the Gaussian action
entirely in terms of $Q$,
\bea
{\cal A}_{\rm G}[Q] &=& 
  - \frac{4}{V}\sum_{\bf k}\sum_{1234}\sum_{rs}\sum_{ij}
  {^i_r(}\delta Q)_{12}({\bf k})\,{^{ij}_{rs}(}A^{-1})_{12,34}({\bf k})
\nonumber\\
&&\times  {^j_s(}\delta Q)_{34}(-{\bf k})
   +{\cal A}_{\rm dis}[\delta Q] + {\cal A}_{\rm int}[\delta Q]\quad,
\label{eq:3.5}
\eea
with $A^{-1}$ the inverse of the matrix $A$ defined in Eq.\ (\ref{eq:3.2b}).
It is convenient to rewrite this result as
\bml
\label{eqs:3.6}
\bea
{\cal A}_{\rm G}[Q] &=& \frac{-4}{V}\sum_{\bf k}\sum_{1234}\sum_{rs}\sum_{ij}
   {^i_r(}\delta Q)_{12}({\bf k})\,{^{ij}_{rs}{\cal M}}_{12,34}({\bf k})\,
\nonumber\\
   &&\times {^j_s(}\delta Q)_{34}(-{\bf k})\quad,
\label{eq:3.6a}
\eea
where
\bea
{\cal M}_{12,34}({\bf k}) &=& \delta_{13}\delta_{24}\left[(A_{12})^{-1}({\bf k})
   -\openone/\pi\,N_{\rm F}\,\tau\right]
\nonumber\\
&& - \delta_{1-2,3-4}\,\delta_{\alpha_1\alpha_2}\,\delta_{\alpha_1\alpha_3}
      \,B\quad,
\label{eq:3.6b}
\eea
with $\openone$ the unit $8\times 8$ matrix, and $B$ a matrix whose elements
are
\be
B_{rs,ij} = -\delta_{rs}\,\delta_{ij}\,2\,T\,\Gamma^{(i)}\quad,
\label{eq:3.6c}
\ee
\eml%
where $\Gamma^{(0)} = -\Gamma^{(s)}$ and $\Gamma^{(1,2,3)} = \Gamma^{(t)}$.

\subsection{Gaussian propagators}
\label{subsec:III.B}

The Gaussian $Q$ propagators are given in terms of the inverse of the matrix
${\cal M}$ defined in Eq.\ (\ref{eq:3.6b}). As in 
Ref.\ \onlinecite{us_fermions} we solve the redundancy problem inherent in
the matrix field theory due to the symmetry relations, Eqs.\ (\ref{eqs:2.5}),
by formulating the theory entirely in terms of matrix elements $Q_{12}$ with
$n_1\geq n_2$. Then we have
\bml
\label{eqs:3.7}
\be
\bigl\langle\, {^i_r(}\delta Q)_{12}({\bf k})\,{^j_s(}\delta Q)_{34}(-{\bf p})
       \bigr\rangle_{\rm G}
   \hskip -4pt = \delta_{{\bf k},{\bf p}}\,\frac{V}{16}\,{^{ij}_{rs}I_{12}}\ 
        {^{ij}_{rs}{\cal M}}^{-1}_{12,34}({\bf k}),
\label{eq:3.7a}
\ee
where
\be
{^{ij}_{rs}I_{12}} = 1 + \delta_{12}\,\left[-1 + J^{ij}_{rs}\right]\quad,
\label{eq:3.7b}
\ee
with
\bea
J^{ij}_{rs} &=& \frac{1}{4}\,\tr\tau_r\tau_s\tau_3^{\dagger}\,\delta_{r0}\,
   \left[ \tr s_is_js_3^{\dagger} + \left({{+\atop+}\atop{+\atop -}}\right)_i\,
      \tr s_js_is_3^{\dagger}\right]
\nonumber\\
&&+ \frac{1}{4}\,\tr\tau_r\tau_s\tau_3^{\dagger}\,\delta_{r3}\,
   \left[ \tr s_is_js_3^{\dagger} - \left({{+\atop+}\atop{+\atop -}}\right)_i\,
      \tr s_js_is_3^{\dagger}\right]
\nonumber\\
&&\hskip 30pt + 2\,\delta_{rs}\,\delta_{ij}\,\left[\delta_{r0}\,\delta_{i0}
   + \delta_{r3}\,(1-\delta_{i0})\right]\quad.
\label{eq:3.7c}
\eea
\eml%

Here and in the following, $\langle\ldots\rangle_{\rm G}$ denotes an average
taken with the Gaussian action. To determine the inverse of ${\cal M}$ we 
notice that the only nonzero matrix elements of ${^{ij}_{rs}A}$, 
Eq.\ (\ref{eq:3.2b}), are the diagonal and the antidiagonal
elements in our $\tau\otimes s$ basis, and that the same is true for the 
inverse of $A$. We find
\bml
\label{eqs:3.8}
\bea
{^{ij}_{rs}{\cal M}}^{-1}_{12,34}({\bf k}) &=& \delta_{13}\,\delta_{24}\,
 {^{ij}_{rs}C}_{n_1n_2}({\bf k}) + \delta_{1-2,3-4}\,\delta_{\alpha_1\alpha_2}\,
   \delta_{\alpha_1\alpha_3}\,
\nonumber\\
&&\qquad\qquad\quad\times{^{ij}_{rs}E}_{n_1n_2,n_3n_4}({\bf k})\quad,
\label{eq:3.8a}
\eea
where
\bea
E_{n_1n_2,n_3n_4}({\bf k}) &=& C_{n_1n_2}({\bf k})\,B\,
\nonumber\\
&&\hskip -60pt \times \left[\openone
 - \sum_{n_5n_6}\delta_{n_1-n_2,n_5-n_6}\,C_{n_5n_6}({\bf k})\,B\right]^{-1}\,
          C_{n_3n_4}({\bf k}) \quad.
\nonumber\\
\label{eq:3.8b}
\eea
Here $B$ is the matrix given by Eq.\ (\ref{eq:3.6c}), and $C$ is defined as
\bea
C_{nm}({\bf k}) &=& (m^{00} + m^{33})\,({\cal D}_{nm}^{+}({\bf k}) 
                                         + {\cal D}_{nm}^{-}({\bf k}))/4
\nonumber\\
       && +(m^{00} - m^{33})\,({\cal E}_{nm}^{+}({\bf k}) 
                                         + {\cal E}_{nm}^{-}({\bf k}))/4
\nonumber\\
       && +(m^{03} + m^{30})\,({\cal D}_{nm}^{+}({\bf k})
                                         - {\cal D}_{nm}^{-}({\bf k}))/4
\nonumber\\
       && +(m^{03} - m^{30})\,({\cal E}_{nm}^{+}({\bf k})
                                         - {\cal E}_{nm}^{-}({\bf k}))/4
\label{eq:3.8c}
\eea
Here we have defined
\bea
{\cal D}_{nm}^{\pm}({\bf k}) &=& \phi_{nm}^{\pm}({\bf k})/
                     [1-\phi_{nm}^{\pm}({\bf k})/\pi N_{\rm F}\tau]\quad,
\label{eq:3.8d}\\
{\cal E}_{nm}^{\pm}({\bf k}) &=& \eta_{nm}^{\pm}({\bf k})/
                     [1-\eta_{nm}^{\pm}({\bf k})/\pi N_{\rm F}\tau]\quad,
\label{eq:3.8e}
\eea
with
\bea
\phi_{nm}^{\pm}({\bf k})&=&\varphi_{nm}^{+}({\bf k}) \pm \psi_{nm}^{+}({\bf k})
   \quad,
\label{eq:3.8f}\\
\eta_{nm}^{\pm}({\bf k})&=&\varphi_{nm}^{-}({\bf k}) \pm \psi_{nm}^{-}({\bf k})
   \quad,
\label{eq:3.8g}
\eea
in terms of
\bea
\varphi_{nm}^{\pm}({\bf k}) &=& \varphi_{nm}^{00}({\bf k}) \pm
                              \varphi_{nm}^{33}({\bf k})\quad,
\label{eq:3.8h}\\
\psi^{\pm}_{nm}({\bf k}) &=& \varphi^{03}_{nm}({\bf k}) \pm
                           \varphi^{30}_{nm}({\bf k})\quad,
\label{eq:3.8i}
\eea
\eml%

Let us discuss these results. First of all, we notice that setting the
magnetization equal to zero results in ${\cal D}^+ = {\cal D}^- = {\cal E}^+
= {\cal E}^- \equiv {\cal D}$, and we recover the results of 
Ref.\ \onlinecite{us_fermions}. In particular, ${\cal D}_{nm}$ is diffusive
for $nm<0$, and hence the spin-singlet and spin-triplet channels of the
Gaussian propagator $C$ are all soft. In the magnetic case, this changes.
${\cal D}^+$ and ${\cal D}^-$ are still soft, 
and given by the nonmagnetic result with the Fermi energy shifted by 
$\pm\Delta$, respectively,
\be
{\cal D}^{\pm}_{nm}({\bf k}) = \frac{\pi N_{\rm F}^{\pm}}{D^{\pm}{\bf k}^2
   + \vert\Omega_{n-m}\vert}\quad.
\label{eq:3.9}
\ee
This holds for $nm<0$, and in the limit of small frequencies and wavenumbers.
For $nm>0$, ${\cal D}^{\pm}$ is finite in that limit. $N_{\rm F}^{\pm}$ and
$D^{\pm}$ are the density of states at the Fermi level, and the Boltzmann
diffusivity, respectively, of an electron system whose Fermi energy has been
shifted by $\pm\Delta$. The spin-singlet, and the longitudinal component of the
spin-triplet, are thus still diffusive, as one would expect. However,
the transverse component of the spin-triplet is massive. A calculation
yields, for $nm<0$,
\bml
\label{eqs:3.10}
\be
{\cal E}_{nm}^{\pm}({\bf k}) = \frac{\pi N_{\rm F}^{\pm}}
   {\left(\frac{1}{\tau^{\pm}} - \frac{1}{\tau}\right)
       + (\vert\Omega_{n-m}\vert + D^{\pm}{\bf k}^2)\frac{\tau^{\pm}}{\tau}}
            \quad,
\label{eq:3.10a}
\ee
where we have defined
\be
\tau^{\pm} = \frac{\tau}{1 \pm 2i\Delta\tau}\quad.
\label{eq:3.10b}
\ee
\eml%
The mass of these ``spin-diffusons'' is thus proportional to the magnetization.
However, the ``interacting'' part $E$ of the propagator, Eqs.\ (\ref{eqs:3.8}),
is still soft even in the transverse spin-triplet channels. To see this,
consider the saddle-point equations (\ref{eqs:2.14}). Multiplying the first
of these equations by ${\cal F}_n({\bf k})$ and the second one by
${\cal G}_n({\bf k})$, subtracting the second equation from the first one
and integrating over the wavevector, we obtain $\varphi_{nn}^{\pm}({\bf k}=0)
= 2iF_n/\Delta_n$ with $F_n$ from Eq.\ (\ref{eq:2.12b}). Using this, it
is easy to show that
\be
T\sum_n {\cal E}_{nn}^{\pm}({\bf k}=0) = -1/2\Gamma^{(t)}\quad.
\label{eq:3.11}
\ee
Therefore, ${^{ij}_{rs}E}$ is still massless for $i,j=1,2$.
These are of course the magnetic Goldstone modes, and the above calculation
just proves that our Gaussian approximation is conserving in the sense that
it correctly reflects the symmetries of the problem, and the resulting soft
modes. We note that there is no frequency restriction on the softness of
the Goldstone modes, $E_{12,34}$ is massless both for $n_1n_2<0$ and for
$n_1n_2>0$. From Eqs.\ (\ref{eqs:3.8}), (\ref{eqs:3.10}), and (\ref{eq:3.11})
we see that the structure of the Goldstone modes is
\bml
\label{eqs:3.12}
\bea
g^{\pm}({\bf p},\Omega_n)&=&\frac{1}{1 + 2\Gamma^{(t)}T\sum_{n_3,n_4}
   \delta_{n_3-n_4,n}\, {\cal E}_{n_3n_4}^{\pm\,\alpha\alpha}({\bf p})}
\nonumber\\
&=&\frac{d}{\pm i\Omega_n - c{\bf p}^2}\quad,
\label{eq:3.12a}
\eea
with $c$ and $d$ constants. The second equality in Eq.\ (\ref{eq:3.12a}) holds
in the limit of small wavenumbers and frequencies. 
The magnetization and disorder dependence of
$c$ and $d$ is quite complicated, and we write down only the clean limit
results to leading order for small values of $\Delta$, where one finds
\be 
c=\Delta/6k_{\rm F}^2\quad,\quad d=-\Delta/2N_{\rm F}\Gamma^{(t)}\quad.
\label{eq:3.12b}
\ee
\eml%

For later reference we also determine the Gaussian ${\bar\Lambda}$ propagator.
Using Eq.\ (\ref{eq:3.4}) in (\ref{eq:3.2a}), we find
\be
\bigl\langle\,{^i_r(}\delta{\bar\Lambda})_{12}({\bf k})\,
   {^j_s(}\delta{\bar\Lambda})_{34}(-{\bf p})\bigr\rangle_{\rm G}
   \hskip -2pt = \delta_{{\bf k},{\bf p}}\,\frac{-1}{16}\,{^{ij}_{rs}I_{12}}\ 
      {^{ij}_{rs}A}_{12,34} ({\bf k})\ .
\label{eq:3.13}
\ee
Notice that the $\varphi^{ij}_{12}({\bf k})$, Eq.\ (\ref{eq:3.3a}),
and therefore the matrix $A$, Eq.\ (\ref{eq:3.2a}), reduce 
to finite numbers in the limit of low frequencies and small wavenumbers.
The ${\bar\Lambda}$ propagator is thus massive.

Finally, we mention that if one keeps the particle-particle or Cooper 
channel, the corresponding propagators have a mass proportional to the 
magnetization, just like the Cooperons for nonmagnetic electrons in an
external magnectic field are massive.\cite{R} Since we are interested in
universal phenomena that are due to the soft modes in the system,
this justifies our having neglected the Cooper channel.

\subsection{Physical correlation functions}
\label{subsec:III.C}

We now use the results of the preceding subsections to calculate 
some correlation functions of physical interest. Let us start with the
single-particle density of states (DOS) in saddle-point approximation. 
From Eqs.\ (\ref{eq:2.7b},\ref{eq:2.11a},\ref{eq:2.12a},\ref{eq:2.13c}) 
we find for the DOS as a function of imaginary frequency
\be
N(i\omega_n) = \frac{1}{2}\,\left[N_{\rm HF}(\mu+\Delta_n,i\omega_n) + 
                     N_{\rm HF}(\mu-\Delta_n,i\omega_n)\right]\quad.
\label{eq:3.14}
\ee
Here $N_{\rm HF}(\mu,i\omega_n)$ is the DOS for {\em nonmagnetic} electrons
with chemical potential $\mu$ in the disordered Hartree-Fock approximation
of Ref.\ \onlinecite{us_fermions}. In particular, 
$N_{\rm F} = N_{\rm HF}(\mu,i0)$. The DOS as a function of the real
frequency is obtained as 
$N(\omega) = {\rm Im}\,N(i\omega_n\rightarrow\omega+i0)$. We recognize
in Eq.\ (\ref{eq:3.14}) the usual result of Stoner theory. The physical
interpretation is that the band gets split into a band of up-spin-electrons
and a band of down-spin-electrons, with the energy splitting proportional
to the magnetization.\cite{Moriya}

Also of interest are the density susceptibility, and the longitudinal and
transverse spin susceptibilities. Let us define a general susceptibility
\bea
{^{ij}_{rs}\chi}({\bf q},\Omega_n) &=& 16\,T\sum_{m_1,m_2}\bigl\langle\,
   {^i_r(}\delta Q)_{m_1-n,m_1}^{\alpha\alpha}({\bf q})\,
\nonumber\\
   &&\times {^j_s(}\delta Q)_{m_2,m_2+n}^{\alpha\alpha}(-{\bf q})\bigr\rangle
       \quad.
\label{eq:3.15}
\eea
By adding an appropriate source term, or by writing the density susceptibility
$\chi_n$ as a fermionic expectation value and translating to a $Q$-field 
correlation by means of Eq.\ (\ref{eq:2.2}), we find
\bml
\label{eqs:3.16}
\bea
\chi_{\rm n}({\bf q},\Omega_n) &=& {^{00}_{00}\chi}({\bf q},\Omega_n)
   + {^{00}_{33}\chi}({\bf q},\Omega_n)
   -i\,{^{00}_{03}\chi}({\bf q},\Omega_n)
\nonumber\\
  && -i\,{^{00}_{30}\chi}({\bf q},\Omega_n)\quad.
\label{eq:3.16a}
\eea
Similarly, one obtains the longitudinal (L) and transverse (T) spin
susceptibilities as
\bea
\chi_{\rm L}({\bf q},\Omega_n) &=& {^{11}_{00}\chi}({\bf q},\Omega_n)
   + {^{11}_{33}\chi}({\bf q},\Omega_n)
   -i\,{^{11}_{03}\chi}({\bf q},\Omega_n)
\nonumber\\
  && -i\,{^{11}_{30}\chi}({\bf q},\Omega_n)\quad.
\label{eq:3.16b}
\eea
\bea
\chi_{\rm T}({\bf q},\Omega_n) &=& {^{33}_{00}\chi}({\bf q},\Omega_n)
   + {^{33}_{33}\chi}({\bf q},\Omega_n)
   -i\,{^{33}_{03}\chi}({\bf q},\Omega_n)
\nonumber\\
  && -i\,{^{33}_{30}\chi}({\bf q},\Omega_n)\quad.
\label{eq:3.16c}
\eea
\eml%

We can calculate these susceptibilities explicitly in Gaussian approximation
by using the results of Sec.\ \ref{subsec:III.B}. For the density
susceptibility we find
\bml
\label{eqs:3.17}
\be
\chi_{\rm n} = \frac{\chi_+ + \chi_- + 4\Gamma^{(t)}\,\chi_+\,\chi_-}
   {1 + (\Gamma^{(t)}-\Gamma^{(s)})(\chi_+ + \chi_-) 
   - 4\Gamma^{(s)}\Gamma^{(t)}\chi_+\,\chi_-}\quad.
\label{eq:3.17a}
\ee
Here all susceptibilities are understood to be functions of ${\bf q}$ and
$\Omega_n$, and the
\be
\chi_{\pm}({\bf q},\Omega_n) = T\sum_m {\cal D}^{\pm}_{m+n,m}
    ({\bf q}) \quad,
\label{eq:3.17b}
\ee
with ${\cal D}$ from Eq.\ (\ref{eq:3.8d}), are disordered Lindhard functions
for nonmagnetic electrons with the chemical potential shifted by $\pm\Delta_n$.
For small frequencies and wavenumbers, and to leading order in the disorder,
they read
\be
\chi_{\pm}({\bf q},\Omega_n) = \frac{N_{\rm F}D^{\pm}{\bf q}^2}{-i\Omega_n
   + D^{\pm}{\bf q}^2}
\label{eq:3.17c}
\ee
with
\be
D^{\pm} = \frac{2/d}{m_{\rm e}}\,(\mu \pm \Delta)\,\tau\quad,
\label{eq:3.17d}
\ee
\eml%
diffusion constants for electrons in the spin-up and spin-down bands,
respectively, in $d$ spatial dimensions. 
Note that our Gaussian approximation respects particle
number conservation, as expressed by the fact that
$\chi_{\rm n}({\bf q}\rightarrow 0,\Omega_n)\rightarrow 0$. However,
the structure of $\chi_{\rm n}$ is {\em not} diffusive. Rather, in
the limit of small ${\bf q}$ and $\Omega_n$ it takes the form
\be
N_{\rm F}\ \frac{-i\Omega_n{\bf q}^2a_1 + {\bf q}^4a_2}{(-i\Omega_n)^2
   -i\Omega_n{\bf q}^2a_3 + {\bf q}^4a_4}\quad,
\label{eq:3.18}
\ee
with $a_1$, $a_2$, $a_3$, and $a_4$ four independent parameters that depend on
$D^+$, $D^-$, $N_{\rm F}\Gamma^{(s)}$, and $\Gamma^{(t)}$.

From $\chi_{\rm n}$ we also obtain the conductivity through the identity
\be
\sigma(i\Omega_n) = \lim_{{\bf q}\rightarrow 0}\frac{i\Omega_n}{{\bf q}^2}\,
   \chi_{\rm n}({\bf q},i\Omega_n)\quad.
\label{eq:3.19}
\ee
In the static limit we obtain $\sigma(0) = \sigma_{\rm B}$, where
\be
\sigma_{\rm B} = \frac{N_{\rm F}}{2}\,(D^+ + D^-) = N_{\rm F}\,D\quad,
\label{eq:3.20}
\ee
with $D$ the Boltzmann diffusion constant of a nonmagnetic system with
unshifted chemical potential. In Gaussian approximation, the conductivity
of our itinerant ferromagnet has thus the ordinary Boltzmann value.

The longitudinal spin susceptibility can also be expressed in terms of
$\chi_+$ and $\chi_-$. From Eqs.\ (\ref{eq:3.16b},\ref{eq:3.15},{\ref{eqs:3.7},
\ref{eqs:3.8}) we obtain
\be
\chi_{\rm L} = \frac{\chi_+ + \chi_- - 4\Gamma^{(s)}\chi_+\chi_-}
               {1 + (\Gamma^{(t)}-\Gamma^{(s)})(\chi_+ + \chi_-)
                  - 4\Gamma^{(s)}\Gamma^{(t)}\chi_+\chi_-}\quad.
\label{eq:3.21}
\ee
We note again that $\chi_{\rm L}$, like $\chi_{\rm n}$, is massless, but
not diffusive.
In the clean limit, $\tau\rightarrow\infty$, $\chi_+$ and $\chi_-$ become
Lindhard functions proper with shifted chemical potentials. In that limit
we recover the RPA result of Izuyama et al.,\ \cite{IKK} if we take into
account that their Hubbard model with coupling constant $v$ corresponds to
the special case $\Gamma^{(t)} = \Gamma^{(s)} = -v/2$ in our notation. Another
special case is the nonmagnetic one, where $\chi_+ = \chi_- = \chi_0$,
with $\chi_0$ the (disordered) Lindhard function with chemical potential
$\mu$. In that case, Eq.\ (\ref{eq:3.21}) reduces to the ordinary RPA
result, $\chi_{\rm L} = 2\chi_0/(1 + 2\Gamma^{(t)}\chi_0)$.

Finally, we calculate the transverse spin susceptibility. We find
\bea
\chi_{\rm T}({\bf q},\Omega_n)&=&
   \frac{T\sum_m {\cal E}^{+}_{m+n,m}({\bf q})}
   {1 + 2\Gamma^{(t)}T\sum_m{\cal E}^{+}_{m+n,m}({\bf q})}
\nonumber\\
&&+ \frac{T\sum_m {\cal E}^{-}_{m+n,m}({\bf q})}
   {1 + 2\Gamma^{(t)}T\sum_m{\cal E}^{-}_{m+n,m}({\bf q})}\quad.
\label{eq:3.22}
\eea
with ${\cal E}^{\pm}$ from Eq.\ (\ref{eq:3.8d}).

To discuss this result, we use again Eq.\ (\ref{eq:3.11}). Inserting this
into Eq.\ (\ref{eq:3.22}) yields
$1/\chi_{\rm T}({\bf q}=0,\Omega_n=0) = 0$, which is indicative of the 
Goldstone modes. An expansion in powers of ${\bf q}$ and $\Omega_n$
yields the familiar quadratic dispersion relation for 
magnons in itinerant ferromagnets, viz.
\be
\chi_{\rm T}({\bf q},i\Omega_n) = \frac{-1}{\Gamma^{(t)}}\ \left[ 
    g^{+}({\bf q},i\Omega_n) + g^{-}({\bf q},i\Omega_n)\right]\quad,
\label{eq:3.23}
\ee
with $g^{\pm}$ from Eq.\ (\ref{eq:3.12a}). In the limit
$\vert{\bf k}\vert/k_{\rm F},\Omega_n/\epsilon_{\rm F} << 
\Delta/\epsilon_{\rm F} << 1$, this result agrees with the 
one obtained with elementary methods.\cite{Moriya}

\section{Effects of the Goldstone modes in perturbation theory}
\label{sec:IV}

In this section we examine the contributions of the ferromagnetic Goldstone  
modes (FMGM) or spin waves
discussed in the previous subsection to the DOS, 
$N(\omega)$, and the electrical conductivity, $\sigma(\Omega)$. 
In particular, we focus on the contributions of these soft modes to the
nonanalytic frequency dependences of these quantities. Our goal is
to determine whether or not these additional, compared to a Fermi liquid
state, soft modes that are due to the long-range ferromagnetic order,
contribute terms to $N(\omega)$ and $\sigma (\Omega)$ that are as strong
as the usual weak localization effects.\cite{LeeRama} 
We will show they do in the case of the conductivity in three-dimensions,
but not in two-dimensions, and not in the case of the DOS in any dimension.
This strongly suggests
that the FMGM are not important, at least near two-dimensions, in
determining the properties of the metal insulator transition from a
ferromagnetic metal to a ferromagnetic insulator. This last point will
be important in (II).

\subsection{Single-particle density of states}
\label{subsec:IV.A}

The computation of the DOS is straightforward using Eq.\ (\ref{eq:2.7b}) 
as a starting point. 
We first write the action, Eqs.\ (\ref{eqs:2.8}), as the saddle-point
contribution, plus the Gaussian part, Eq.\ (\ref{eq:3.1}),
which is quadratic in $Q$ and ${\wt\Lambda}$, plus non-Gaussian (cubic and
higher order) terms,
\bml
\label{eqs:4.1}
\be
{\cal A}[Q,{\wt\Lambda}] = {\cal A}[Q_{\rm sp},{\wt\Lambda}_{\rm sp}]
    + {\cal A}_{\rm G}[\delta Q,\delta{\wt\Lambda}]
    + \sum_{\ell=3}^{\infty }{\cal A}_{\ell}[\delta{\wt\Lambda}]\quad,
\label{eq:4.1a}
\ee
where
\be
{\cal A}_{\ell }[\delta{\wt\Lambda}] = -\frac{1}{2\ell}\,
     \Tr (iG_{\rm sp}\delta{\wt\Lambda})^{\ell} \quad.
\label{eq:4.1b}
\ee
By introducing $\delta{\bar\Lambda}$ again as defined in Eq.\ (\ref{eq:3.4}),
we can decouple $\delta Q$ and $\delta{\bar\Lambda}$ in the Gaussian term,
at the expense of having the higher order terms depend on both $Q$ and
${\bar\Lambda}$,
\be
{\cal A}[Q,{\bar\Lambda}] = {\cal A}_{\rm sp}
   + {\cal A}_{\rm G}[\delta Q,\delta{\bar\Lambda}]
    + \sum_{\ell=3}^{\infty }{\cal A}_{\ell}[2A^{-1}(\delta{\bar\Lambda}
        - \delta Q)]\ .
\label{eq:4.1c}
\ee
\eml%

We proceed by writing the one-point
$Q$-correlation function on the right hand side of 
Eq.\ (\ref{eq:2.7b}) as,
\be
\bigl\langle\,{_0^0Q}_{nn}^{\alpha \alpha }({\bf x})\bigr\rangle = 
   {^0_0Q}_{nn}^{\alpha\alpha}\bigl\vert_{\rm sp} 
   + \bigl\langle\,{^0_0(}\delta Q)_{nn}^{\alpha\alpha}\bigr\rangle\quad.
\label{eq:4.2}
%\tag{2.38}
\ee
The diagrammatic loop expansion for the irreducible part of 
$\langle\delta Q\rangle$, or equivalently for the one-point vertex function,
is shown in Fig.\ \ref{fig:4.1}. 
\begin{figure}[tb]
%\vskip 2.5cm
\epsfxsize=6.25cm
\centerline{\epsffile{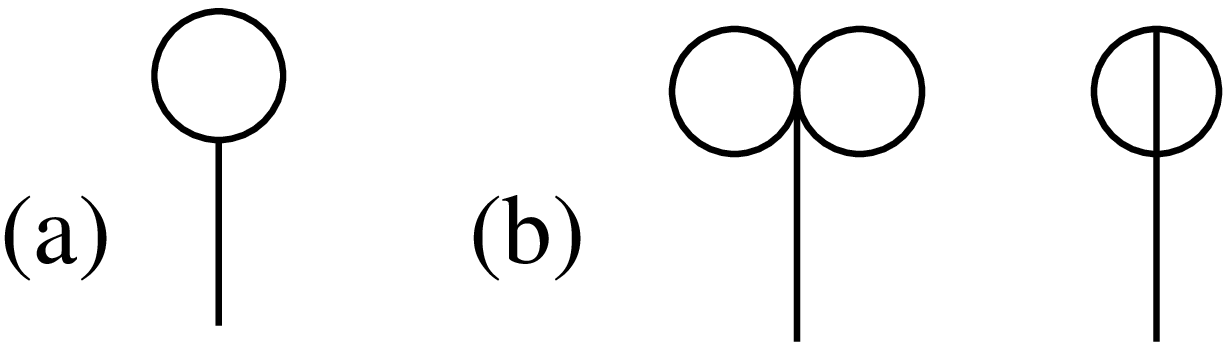}}
\vskip 0.5cm
\caption{One-loop (a) and two-loop (b) contributions to the one-point
 vertex function.} 
\label{fig:4.1}
\end{figure}
The one-loop term, Fig.\ \ref{fig:4.1}(a), is given analytically by
\be
\langle \delta Q_{nn}^{\alpha\alpha}({\bf x})\rangle_{\rm 1-loop} = 
   \bigl\langle\delta Q_{nn}^{\alpha\alpha}({\bf x})\,
     {\cal A}_{3}[2A^{-1}(\delta{\bar\Lambda}
        - \delta Q)]\,\bigr\rangle_{\rm G} \quad.  
\label{eq:4.3}
%\tag{2.39}
\ee
Equation\ (\ref{eq:4.3}) is evaluated by using Wick's theorem and the
Gaussian propagators, Eqs.\ (\ref{eqs:3.7}) and (\ref{eq:3.13}).
The leading nonanalyticities are given by contributions where the loop in
Fig.\ \ref{fig:4.1}(a) is a soft mode. That is, the loop must be a
$Q$-propagator, since the ${\bar\Lambda}$-propagator is massive,
see Eq.\ (\ref{eq:3.13}). As we have
seen in Sec.\ \ref{sec:III}, there are two different types of soft 
$Q$-propagators, viz. the diffusive modes or diffusons in the spin-singlet 
and longitudinal spin-triplet channels, and the Goldstone modes or
spin waves in the transverse spin-triplet channels. The former,
Eq.\ (\ref{eq:3.9}), contribute a term that is very similar to the weak 
localization contribution to the DOS for nonmagnetic 
electrons in an external magnetic field. We will discuss this contribution 
in detail in (II). For now we 
just consider the functional form of these contributions. They take the
form of a frequency-momentum integral over a diffuson propagator. The
most divergent part of the latter is the interacting piece, which is
denoted by $E$ in Eq.\ (\ref{eq:3.8b}). Inspection shows that this piece
is essentially a diffusion propagator squared, i.e. it scales
like $1/\omega^2 \sim 1/{\bf k}^4$ (here, and in similar arguments below,
${\bf k}$ and $\omega$ denote generic internal wavevectors and frequencies.) 
Dimensional analysis then yields
\bml
\label{eqs:4.4}
\be
%\bigl\langle \delta Q_{nn}^{\alpha\alpha}({\bf x})\bigr\rangle
N(\mu+\omega)_{\rm 1-loop,diffusons} 
     = O(\omega^{(d-2)/2})\quad.
\label{eq:4.4a}
\ee
The other soft mode contribution is due to the FMGM, or spin waves. They
take the form of a frequency-momentum integral over the FMGM propagator,
Eq.\ (\ref{eq:3.12a}), which scales like 
$1/{\bf k}^2 \sim 1/\omega$. We thus have
\be
%\bigl\langle \delta Q_{nn}^{\alpha\alpha}({\bf x})\bigr\rangle
N(\mu+\omega)_{\rm 1-loop,FMGM} 
   = O(\omega^{d/2}) \quad.
\label{eq:4.4b}
\ee
\eml%

We conclude that the FMGM contributions to the DOS are subleading compared
to the diffuson contributions. We will discuss this result further in
Sec.\ \ref{sec:V} below. Here we just mention that it implies that the
FMGMs are irrelevant, as far as the DOS is concerned,
for all leading universal effects in the limit of
long wavelengths and low frequencies, in particular for the critical
behavior at the metal-insulator transition from a FM metal to a FM
insulator near $d=2$.

\subsection{Electrical conductivity}
\label{subsec:IV.B}

We now investigate the same point for the electrical conductivity.
In terms of Grassmann variables, the dynamical electrical conductivity 
as a function of the real frequency $\Omega$ is given by the Kubo
formula,\cite{Kubo,Mahan}
\bea
\sigma (\Omega) &=& \frac{in/m_{\rm e}}{\Omega }
    + \frac{iT}{\Omega V m_{\rm e}^2}
\sum_{n_{1}n_{2}}\sum_{\sigma \sigma'}\int d{\bf x}\,d{\bf x}'
\nonumber\\
&&\times\langle{\bar\psi}_{n_1\sigma}^{\alpha}({\bf x})\,\partial_{x_1}\,
   \psi_{n_1+m,\sigma}^{\alpha}({\bf x})
      {\bar\psi}_{n_2\sigma'}^{\alpha}({\bf x}')\,
\nonumber\\
&&\times\partial_{x_1'}\,\psi_{n_2-m,\sigma'}^{\alpha}({\bf x}')\rangle
     \bigl\vert_{i\Omega_m\rightarrow\Omega +i0}\quad.
\label{eq:4.5}
%\tag{2.41}
\eea
Here ${\bf x} = (x_1,x_2,x_3)$, etc.
The gradient operators in Eq.\ (\ref{eq:4.5}) imply that $\sigma(\Omega)$ 
cannot be written as a simple $Q$-correlation function. At this point we have
two choices. We can either generalize the fields we work with, or, we
can introduce a suitable source field to generate $\sigma$. We choose
the second path, following Refs.\ \onlinecite{CastellaniKotliar,cds_trk}.

\subsubsection{Source formalism}
\label{subsubsec:IV.B.1}

To the fermionic action, Eq.\ (\ref{eq:2.1b}), we  add a source term $S_j$,
\bml
\label{eqs:4.6}
\be
S_j = - \sum_{\alpha}\sum_m j_m^{\alpha} \int d{\bf x} \sum_n\sum_{\sigma}
   {\bar\psi}_{n\sigma}^{\alpha}({\bf x})\,\partial_{x_1}\,\psi_{n-m,\sigma}
    ^{\alpha}({\bf x})\quad.
\label{eq:4.6a}
\ee
After integrating out the Grassmann fields, the action 
${\cal A}[Q,{\wt\Lambda},j]$ is given 
by Eq.\ (\ref{eq:2.8a}) with the $\Tr\ln$ term replaced by,
\be
{\cal A}_{\ln }[{\wt\Lambda},j] = \frac{1}{2}\,
   \Tr\ln (G_0^{-1} + JL - i{\wt\Lambda}) \quad,  
\label{eq:4.6b}
%\tag{2.42a}
\ee
with
\be
(JL)_{12} = \frac{1}{2}\sum_{m} j_m^{\alpha_1}L_{12,m}\partial_{x_1}\quad,  
\label{eq:4.6c}
%\tag{2.42b}
\ee
where
\be
L_{12,m} = \delta_{\alpha_1\alpha_2}\bigl[(\tau_{-}\!\otimes s_0)\,
   \delta_{n_1,n_2+m} - (\tau_{+}\!\otimes s_0)\,\delta_{n_2,n_1+m}\bigr],
\label{eq:4.6d}
%\tag{2.42c}
\ee
with 
\be
\tau_{\pm} = \tau_0 \pm i\tau_3\quad.
\label{eq:4.6e}
\ee
\eml%
In the presence of the source, the partition function, Eq.\ (\ref{eq:2.3}),
turns into a $j$-dependent generating functional,
\bml
\label{eqs:4.7}
\be
Z[j] = \int D[Q]\int D[{\wt\Lambda}]\,e^{{\cal A}[Q,{\wt\Lambda},j]}\quad.
\label{eq:4.7a}
\ee
In terms of derivatives of $Z[j]$, $\sigma (\Omega)$ is given by,
\be
\sigma (\Omega) = -\frac{in/m_{\rm e}}{\Omega}
           + \frac{iT}{\Omega V m_{\rm e}^2}\,
   \frac{\partial^2}{\partial j_m^{\alpha}\partial j_{-m}^{\alpha}}\,
      \ln Z[j]\Bigl\vert_{j=0\atop i\Omega_m\rightarrow \Omega +i0}\ .  
\label{eq:4.7b}
\ee
\eml%

There are several possible ways to proceed at this point. References
\onlinecite{CastellaniKotliar,cds_trk} derived a nonlinear $\sigma$ model.
Here, we will use instead a simple perturbation expansion. 
We first expand ${\cal A}_{\ln}$,
Eq.\ (\ref{eq:4.6b}), in powers of $j$ and $\delta{\wt\Lambda}$:
\be
{\cal A}_{\ln }[{\wt\Lambda}] = \sum_{i,\ell=0}^{\infty} A_{(i,\ell)}
     \quad,  
\label{eq:4.8}
%\tag{2.44}
\ee
with $A_{(i,\ell)} = O(j^i,\delta{\wt\Lambda}^{\ell})$.
According to Eq.\ (\ref{eq:4.7b}) we can restrict ourselves to terms 
with $i\leq 2$.
The saddle point contribution is,
\be
A_{(2,0)} = -\frac{1}{4}\,\Tr (G_{\rm sp}JLG_{\rm sp}JL)\quad.  
\label{eq:4.9}
%\tag{2.45}
\ee
In this approximation, and using Eqs.\ (\ref{eqs:2.13})
and (\ref{eqs:4.6}) in Eq.\ (\ref{eq:4.9}), and the result in
Eqs.\ (\ref{eqs:4.7}), one obtains the Boltzmann equation result for 
$\sigma$ as given by Eq. (\ref{eq:3.20}).

For the fluctuation corrections to the Boltzmann conductivity, there are two
classes of terms: $A_{(1,\ell)}$ and $A_{(2,\ell)}$, 
both with $\ell\geq 2$.
Since there are two $j$-derivatives in Eq. (\ref{eq:4.7b}) for $\sigma$,
the relevant
correlation functions are $\langle A_{(1,\ell)}^2\rangle_{\rm G}$ and
$\langle A_{(2,\ell)}\rangle_{\rm G}$.
\begin{figure}[tb]
%\vskip 2.5cm
\epsfxsize=8.25cm
\centerline{\epsffile{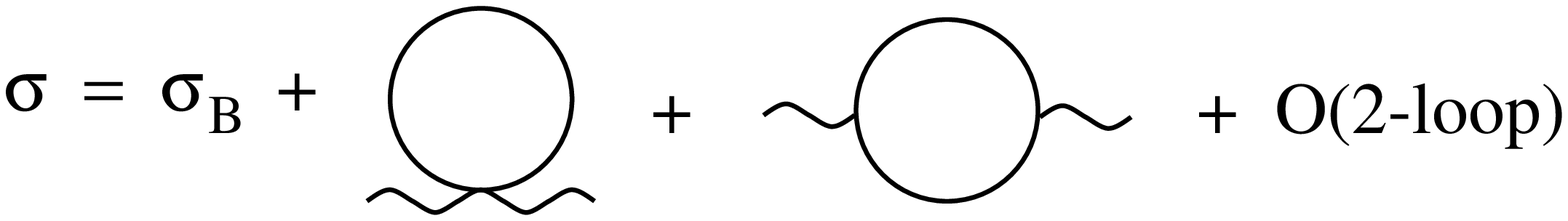}}
\vskip 0.5cm
\caption{Diagrammatic contributions to the conductivity. Wavy lines denote
 sources $JL$, and solid lines denote $\delta{\wt\Lambda}$ propagators.}
\label{fig:4.2}
\end{figure}
The diagrams that give the one-loop contributions to $\sigma$ are shown in
Fig.\ \ref{fig:4.2}. Analytically one needs,
\bml
\label{eqs:4.10}
\be
A_{(1,2)} = -\frac{1}{2}\,\Tr (G_{\rm sp}JLG_{\rm sp}\delta{\wt\Lambda}G_{\rm sp}
          \delta{\wt\Lambda})\quad,
\label{eq:4.10a}
%\tag{2.46a}
\ee
and
\bea
A_{(2,2)} &=& \frac{3}{8}\,\left[\Tr (G_{\rm sp}JLG_{\rm sp}\delta{\wt\Lambda}
            G_{\rm sp}JLG_{\rm sp}\delta{\wt\Lambda})\right.
\nonumber\\
&& \left. +\Tr (G_{\rm sp}JLG_{\rm sp}JLG_{\rm sp}\delta{\wt\Lambda}G_{\rm sp}
    \delta{\wt\Lambda})\right]\quad.
\label{eq:4.10b}
%\tag{2.46b}
\eea
\eml%
Now we write the loop expansion of the conductivity as
\be
\sigma (\Omega\rightarrow 0) = \sigma_{\rm B} + \sum_{i=1}^{\infty} 
      \sigma_{i}(\Omega\rightarrow 0)\quad,
\label{eq:4.11}
%\tag{2.47}
\ee
with $\sigma_i$ the $i^{\rm th}$ term in the loop expansion. To one-loop
order, Eqs. (\ref{eq:4.7b}) and (\ref{eq:4.8}) give,
\bea
\sigma_{1}(\Omega) &=& \frac{iT}{\Omega V m_{\rm e}^2}\,
    \frac{\partial^2}{\partial j_m^{\alpha}\partial j_{-m}^{\alpha }}\,
    \Biggr\vert_{\hskip -25pt j=0\atop i\Omega_m\rightarrow\Omega +i0}
       \hskip -20pt \biggl[\langle A_{2,2}\rangle_{\rm G}
 +\frac{1}{2}\,\langle A_{(1,2)}^{2}\rangle_{\rm G}^c
                \biggr]
\nonumber\\
&\equiv&\left[\sigma_{(1,1)} + \sigma_{(1,2)}
             \right]_{i\Omega_m\rightarrow\Omega + i0}\quad,
\label{eq:4.12}
%\tag{2.48}
\eea
where $\langle\ldots\rangle^c$ denotes that only connected diagrams are 
to be taken into
account. $\sigma_{(1,1)}$ and $\sigma_{(1,2)}$ are defined, in order,
as the two terms in the previous equality. They are graphically
represented by the first and second diagram, respectively, 
in Fig.\ \ref{fig:4.2}.

\subsubsection{Analysis of the one-loop terms}
\label{subsubsec:IV.B.2}

The evaluation of the one-loop terms given by Eq.\ (\ref{eq:4.12}) 
is straightforward but tedious. $\sigma_{(1,1)}$ and $\sigma_{(1,2)}$
are given by two- and four-$\delta{\wt\Lambda}$ correlation functions,
respectively. Due to the coupling between $\delta{\wt\Lambda}$ and
$\delta Q$ these correlation functions contain both diffuson and FMGM 
contributions, and hence are soft.

We first consider
$\sigma_{(1,1)}$. Structurally, this term is analogous to the one-loop
correction to the DOS that was calculated in Sec.\ \ref{subsec:IV.A}.
Therefore, while the diffuson contribution to this diagram leads to
the usual weak-localization correction of order $\Omega^{(d-2)/2}$ 
in $2<d<4$, and $\ln\Omega$ in $d=2$, the FMGM contribution is weaker, 
of order $(\Omega^{d/2})$ in $2<d<4$.

The $\sigma_{(1,2)}$ contribution, which is represented by the second diagram
in Fig.\ \ref{fig:4.2}, is more complicated. From the spin structure of
this diagram it follows that the two propagators that form the loop must
either both be diffusons, or both FMGMs; the contribution that mixes these
two modes vanishes. The diagram with two diffusons yields again a 
weak-localization term of $O(\Omega^{(d-2)/2})$, and will be discussed in (II).
The one with two FMGMs has the following structure. Each mode contributes
a factor of $1/{\bf k}^2\sim 1/\omega$ to the integrand. Each gradient 
operator in $A_{(1,2)}$ leads to a factor of $k_x$. Expanding the
propagators to order $\Omega$ and integrating the resulting
term over frequency and wavenumber then in general leads to an 
$\Omega^{(d-2)/2}$ term in $d$-dimensions. Detailed calculations confirm this
structural argument, see below. The nonanalytic frequency dependence of the 
FMGM contributions to the conductivity is thus as strong as that of the diffuson
contributions. However, we find that the prefactor of the FMGM nonanalyticity
is of $O(1)$ in $d=2+\epsilon$, in contrast to that of the diffuson
nonanalyticity, which is of $O(1/\epsilon)$. While the two sets of soft
modes thus give comparable contributions to the nonanalyticity in the
metallic phase in $d=3$, the FMGMs do not lead to a $\ln\Omega$ term in
$d=2$. As in the case of the DOS, the conclusion is that the FMGMs are 
irrelevant for the description of the metal-insulator transition in 
$d=2+\epsilon$.

For the technical derivation of these results, we start by explicitly writing 
out $A_{(1,2)}$ given by Eq.\ (\ref{eq:4.10a}). Note that
according to the argument given above, the singular terms arise from the
wavenumber and frequency dependencies of the soft modes, and all other 
${\bf k}$ and $\omega $ dependence can be neglected. Using this we can 
localize $A_{(1,2)}$ in space and time and obtain,
\be
A_{(1,2)}\approx \sum_{m,\alpha}A_{(1,2)\,m}^{\alpha}\ j_{m}^{\alpha}
   \quad,
\label{eq:4.13}
%\tag{2.49}
\ee
with,
\bml
\label{eqs:4.14}
\bea
A_{(1,2)\,m}^{\alpha} &=& -\frac{1}{4}\sum_{n_1n_2}\sum_{rs,ij}\frac{1}{V}
\sum_{\bf k}ik_x \sum_{\sigma=\pm}\sigma\ {^{ij}_{rs}N}_{n_1}^{-\sigma} 
\sum_{\beta}
\nonumber\\
&&\hskip -34pt \times {_r^i(}\delta{\wt\Lambda})_{n_1,n_1+n_2}^{\alpha\beta}
   ({\bf k})\ {_s^j(}\delta{\wt\Lambda})_{n_1+n_2,n_1+\sigma m}^{\beta\alpha}
       (-{\bf k}) \quad. 
\label{eq:4.14a}
%\tag{2.50a}
\eea
Here ${\bf k} = (k_x,k_y,k_z)$, and $N$ is the matrix element,
\bea
{^{ij}_{rs}N}_n^{\sigma} &=& \sum_{a,b,c=0,3}\,
     \tr(\tau_a\tau_{\sigma}\tau_b\tau_r\tau_c\tau_s)\ \tr(s_a s_b s_i s_c s_j)
\nonumber\\
&&\hskip -20pt \times\int d{\bf x}\,d{\bf y}\,G_n^a({\bf x})\,\partial_{x_1}\,
     G_n^b({\bf x}-{\bf y})\,y_1\,G_{n}^c({\bf y})\quad,
\label{eq:4.14b}
%\tag{2.50b}
\eea
\eml%
with $G^0$ and $G^3$ from Eq.\ (\ref{eq:3.3a}), and $\tau_{\pm}$ from
Eq.\ (\ref{eq:4.6e}).
From Eqs. (\ref{eq:4.11}) - (\ref{eq:4.14a}), and using the factorization 
property of the Gaussian action, we obtain
\bea
\sigma_{(1,2)} &=& -\frac{iT}{16\Omega V m_{\rm e}^2}\sum_{n_1,n_2,n_1',n_2'}\ 
   \sum_{i,j,i',j'}\ \sum_{r,s,r',s'}\ \sum_{\sigma,\sigma'} \sigma\,\sigma'
\nonumber\\
&&\times {^{ij}_{rs}N}_{n_1}^{-\sigma}\ {^{i'j'}_{r's'}N}_{n_1'}^{-\sigma'}\  
\frac{1}{V^2}\sum_{{\bf k},{\bf k}'} k_x\,k'_x\sum_{\beta\beta'}\
\nonumber\\
&&\hskip -35pt\times\left[\bigl\langle\,{^i_r(}\delta{\wt\Lambda})^{\alpha\beta}
    _{n_1,n_1+n_2}({\bf k})\,
{^{i'}_{r'}(}\delta{\wt\Lambda})^{\alpha\beta'}
    _{n_1',n_1'+n_2'}({\bf k}')\bigr\rangle_{\rm G}\right.
\nonumber\\
&&\hskip -30 pt \times\bigl\langle\,{^j_s(}\delta{\wt\Lambda})^{\beta\alpha}
    _{n_1+n_2,n_1+\sigma m}(-{\bf k})\ {^{j'}_{s'}(}\delta{\wt\Lambda})
    ^{\beta'\alpha}_{n_1'+n_2',n_1'-\sigma' m}(-{\bf k}')\bigr\rangle_{\rm G}
\nonumber\\
&&\hskip -35 pt  +\bigl\langle\,{^i_r(}\delta{\wt\Lambda})^{\alpha\beta}
    _{n_1,n_1+n_2}({\bf k})\ {^{j'}_{s'}(}\delta{\wt\Lambda})^{\beta'\alpha}
    _{n_1'+n_2',n_1-\sigma' m}(-{\bf k}')\bigr\rangle_{\rm G}
\nonumber\\
&&\left.\hskip -30 pt\times\bigl\langle\,{^j_s(}\delta{\wt\Lambda})
         ^{\beta\alpha}
    _{n_1+n_2,n_1+\sigma m}(-{\bf k})\ {^{i'}_{r'}(}\delta{\wt\Lambda})
    ^{\alpha\beta'}_{n_1',n_1'+n_2'}({\bf k}')\bigr\rangle_{\rm G}\right]\ .
\nonumber\\
\label{eq:4.15}
\eea
Note that Eqs.\ (\ref{eq:4.14b}) and (\ref{eq:4.15}) show that the 
FMGMs and diffusons are not coupled
together: If $i$ equals one or two, then $j$ must be one or two;
similarly, if $i$ is zero or three, then $j$ must be zero or three
for a nonzero ${^{ij}N}$. In what follows, we restrict ourselves to
the FMGM contributions, so we set $i,j = (1,2)$ and 
$\beta =\beta'=\alpha$ in Eq.\ (\ref{eq:4.15}).
For later reference we also note the symmetry relations
\bml
\label{eqs:4.16}
\bea
{^{11}_{rs}N_n^{\sigma}} &=& {^{22}_{rs}N_n^{\sigma}}\qquad\hskip 4 pt,\quad 
         {^{12}_{rs}N_n^{\sigma}} = -\,{^{21}_{rs}N_n^{\sigma}}\quad,
\label{eq:4.16a}\\
{^{ij}_{33}N_n^{\sigma}} &=& -\,{^{ij}_{00}N_n^{\sigma}},\quad,\quad 
         {^{ij}_{30}N_n^{\sigma}} = {^{ij}_{03}N_n^{\sigma}}\quad.
\label{eq:4.16b}
%\tag{2.52b}
\eea
\eml%

Next we separate the $\delta{\wt\Lambda}$ flucuations into soft $\delta Q$
fluctuations and massive $\delta{\bar\Lambda}$ fluctuations using 
Eq.\ (\ref{eq:3.4}). To make our procedure more transparent, we will
initially integrate out the massive modes in saddle-point approximation,
i.e. we simply drop the $\delta{\bar\Lambda}$. We will improve on this
in Section \ref{subsubsec:IV.B.3} below. In addition to this approximation, 
we keep only terms that contribute to nonanalyticities of $O(\Omega^{(d-2)/2})$
in $\sigma(\Omega)$. This procedure amounts to the replacement,
in Eq.\ (\ref{eq:4.15}) and for $i,j=1,2$,
\bml
\label{eqs:4.17}
\bea
{^{ij}_{rs}N}_n^{\sigma}&\Rightarrow&4\ {^{ij}_{rs}N}_n^{\sigma}/B_n\quad,
\nonumber\\
\delta{\wt\Lambda}&\Rightarrow&\delta Q\quad,
\label{eq:4.17a}
%\tag{2.53a}
\eea
with
\be
B_n \equiv \varphi^{-}_{nn}({\bf k}=0) 
 = \frac{1}{V}\sum_{\bf p}\left[{\cal G}_n^2({\bf p}) - {\cal F}_n^2({\bf p})
      \right]\quad,
\label{eq:4.17b}
%\tag{2.53b}
\ee
\eml%
from Eq.\ (\ref{eq:3.8h}).
In terms of Q-fluctations, the leading FMGM contribution to $\sigma_{(1,2)}$
thus is
\bea
\sigma_{(1,2)} &=& -\frac{iT}{\Omega V m_{\rm e}^2}\sum_{n_1,n_1',n_2,n_2'}\ 
   \sum_{r,s,r',s'}\ \sum_{i,j,i',j'}\ \sum_{\sigma,\sigma'} \sigma\,\sigma'
\nonumber\\ 
&&\times\frac{{^{ij}_{rs}N}_{n_1}^{-\sigma}}{(B_{n_1})^2}\,
   \frac{^{i'j'}_{r's'}N_{n_1'}^{-\sigma'}}{(B_{n_1'})^2}\,
   \frac{1}{V^2}\sum_{{\bf k},{\bf k}'} k_x\,k'_x
\nonumber\\
&&\hskip -35pt \times\biggl[-\bigl\langle\,{_r^i(}\delta Q)_{n_1,n_1+n_2}
    ^{\alpha\alpha} ({\bf k})\,
       {_{r'}^{i'}(}\delta Q)_{n_1',n_1'+n_2'}^{\alpha\alpha}
     ({\bf k}') \bigr\rangle_{\rm G}
\nonumber\\
&&\hskip -30 pt \times \bigl\langle\,{_s^j(}\delta Q)_{n_1+n_2,n_1+\sigma m}
        ^{\alpha\alpha}(-{\bf k})\ 
{_{s'}^{j'}(}\delta Q)_{n_1'+n_2',n_1'- \sigma' m}^{\alpha\alpha}(-{\bf k'})
     \bigr\rangle_{\rm G}
\nonumber\\
&&\hskip -25pt + \bigl\langle\,{_r^i(}\delta Q)_{n_1,n_1+n_2}^{\alpha\alpha}
                  ({\bf k})\ 
    {_{s'}^{j'}(}\delta Q)_{n_1'+n_2',n_1'-\sigma' m}^{\alpha\alpha}(-{\bf k'})
        \bigr\rangle_{\rm G}
\nonumber\\
&&\hskip -15 pt \times\bigl\langle\,{_s^j(}\delta Q)_{n_1+n_2,n_1+\sigma m}
   ^{\alpha\alpha}(-{\bf k})\ 
   {_{r'}^{i'}(}\delta Q)_{n_1',n_1'+n_2'}^{\alpha\alpha}({\bf k'})
    \bigr\rangle_{\rm G}
     \biggr]\ .
\nonumber\\
\label{eq:4.18}
%\tag{2.54}
\eea
We define
\bml
\label{eqs:4.19}
\bea
\bigl\langle\,{_r^i(}\delta Q)_{n,n+m}^{\alpha\alpha}({\bf k})\ 
   {_s^j(}\delta Q)_{n',n'+m'}^{\alpha\alpha}({\bf k'})\bigr\rangle_{\rm G} &=&
     (2\pi)^d\delta ({\bf k}+{\bf k'})
\nonumber\\
&&\hskip -150pt \times{\cal C}_{rs}^{ij}(n,n+m,{\bf k}\vert n',n'+m')\quad.
\label{eq:4.19a}
%\tag{2.55}
\eea
Notice that we have not yet implemented the frequency restrictions that we
used to define our Gaussian theory in Sec.\ \ref{sec:III}, so the
$C^{ij}_{rs}(n_1,n_2,{\bf k}\vert n_3,n_4)$ are related to the propagators
${\cal M}^{-1}$ of Eq.\ (\ref{eq:3.7a}) by
\bea
C^{ij}_{rs}(n_1,n_2,{\bf k}\vert n_3,n_4)&=&\frac{1}{16}\,\biggl[
   \Theta (n_1-n_2)\,\Theta (n_3-n_4)\,
\nonumber\\
&&\hskip 61pt \times {^{ij}_{rs}{\cal M}}^{-1\ \alpha\alpha,\alpha\alpha}
                                                _{n_1n_2,n_3n_4}({\bf k})
\nonumber\\
&&\hskip -95pt \times \Theta (n_1-n_2)\,\Theta (n_4-n_3)\,
   (-)^s\,\left({{+\atop -}\atop{-\atop -}}\right)_j
 {^{ij}_{rs}{\cal M}}^{-1\ \alpha\alpha,\alpha\alpha}_{n_1n_2,n_4n_3}({\bf k})
\nonumber\\
&&\hskip -95pt \times \Theta (n_2-n_1)\,\Theta (n_3-n_4)\,(-)^r\,
   \left({{+\atop -}\atop{-\atop -}}\right)_i
 {^{ij}_{rs}{\cal M}}^{-1\ \alpha\alpha,\alpha\alpha}_{n_2n_1,n_3n_4}({\bf k})
\nonumber\\
&&\hskip -95pt \times \Theta (n_2-n_1)\,\Theta (n_4-n_3)\,(-)^{r+s}
   \left({{+\atop -}\atop{-\atop -}}\right)_i\,
     \left({{+\atop -}\atop{-\atop -}}\right)_j
\nonumber\\
&&\hskip 10pt \times
{^{ij}_{rs}{\cal M}}^{-1\ \alpha\alpha,\alpha\alpha}_{n_2n_1,n_4n_3}({\bf k})
    \biggr]\quad.
\label{eq:4.19b}
\eea
\eml%
Here we have used the symmetry relations of the $Q$-matrices, 
Eqs.\ (\ref{eqs:2.5}). We further define the quantity $M^{abc}_n$ by 
(cf. Eq.\ (\ref{eq:4.14b}))
\be
M^{abc}_n = \int d{\bf x}\,d{\bf y}\,G_n^a({\bf x})\,\partial_{x_1}\,
   G_n^b({\bf x}-{\bf y})\,y_1 \,G_n^c({\bf y})\quad. 
\label{eq:4.20}
%\tag{2.56}
\ee
Carrying out the spin and quaternion sums in Eq.\ (\ref{eq:4.18}), 
and using Eqs.\ (\ref{eqs:4.16}), (\ref{eqs:4.19}), and (\ref{eq:4.20}),
one obtains
\bml
\label{eqs:4.21}
\bea
\sigma_{(1,2)} &=& \frac{-16T}{\Omega m_{\rm e}^2}\sum_{n_1,n_2,n_1',n_2'}\ 
   \frac{1}{V}\sum_{\bf k}
   \frac{k_x^2}{B_{n_1}^2\, B_{n'_1}^2}\sum_{\sigma\sigma'}\sigma\,\sigma'
\nonumber\\
&&\hskip -20pt \times\biggl\{N_{n_1}^{(1)}\,N_{n_1'}^{(1)}\,
      \Bigl[\bigl(\sigma{\cal C}_{00}^{11} - \sigma'{\cal C}_{33}^{11}\bigr)\,
           \bigl(\sigma'{\cal C}_{00}^{11} - \sigma{\cal C}_{33}^{11}\bigr)
\nonumber\\
&&\hskip 20pt + \bigl(\sigma{\cal C}_{03}^{12} + \sigma'{\cal C}_{30}^{12}\bigr)
        \bigl(\sigma'{\cal C}_{03}^{21} + \sigma{\cal C}_{30}^{21}\bigr)\Bigr]
\nonumber\\
&&\hskip -15pt + N_{n_1}^{(2)}\,N_{n_1'}^{(1)}\,
   \Bigl[\bigl(\sigma{\cal C}_{30}^{12} + \sigma'{\cal C}_{03}^{12}\bigr)\,
        \bigl(\sigma'{\cal C}_{00}^{11} - \sigma{\cal C}_{33}^{11}\bigr)
\nonumber\\
&& - \bigl(\sigma{\cal C}_{00}^{11} - \sigma'{\cal C}_{33}^{11}\bigr)
      \bigl(\sigma'{\cal C}_{30}^{12} 
         + \sigma{\cal C}_{03}^{12}\bigr)\Bigr] \biggr\}\quad,
\nonumber\\
\label{eq:4.21a}
% \tag{2.57}
\eea
where the frequency labels on the ${\cal C}$ are
${\cal C}(n_1,n_1+n_2,{\bf k}\vert n_1',n_1'+n_2')$ for the first
bracket in each of the four terms, and 
${\cal C}(n_1+n_2,n_1+\sigma m,-{\bf k}\vert n_1'+n_2',n_1'-\sigma' m)$ 
for the second bracket, and
\bea
N^{(1)}_n&=&M^{333}_n - M^{300}_n - M^{030}_n + M^{003}_n\quad.
\label{eq:4.21b}\\
N^{(2)}_n&=& M^{033}_n + M^{303}_n - M^{330}_n - M^{000}_n\quad,
\label{eq:4.21c}
\eea
\eml%
In writing Eqs.\ (\ref{eq:4.21a}) we have rearranged the frequency labels
and made approximations that do not affect the leading nonanalytic terms
we are aiming to calculate. In particular, we have dropped dependences of
the prefactors $B_n$ and $N^{(1,2)}_n$ on the ``hydrodynamic'' frequency
labels $n_2$ and $m$.

We next perform the sums over $\sigma$ and $\sigma'$ in Eq.\ (\ref{eq:4.21a}).
It turns out that none of the terms that involve 
$C^{12}$ or $C^{21}$ contribute
to the leading nonanalytic frequency dependence of the conductivity. We are
thus left with only one term, which depends on ${^{11}_{00}{\cal M}^{-1}}$ 
only. Keeping again only leading terms, the latter
can be expressed in terms of the fundamental Goldstone propagator
\bea
g({\bf k},i\Omega_n)&=&g^+({\bf k},\Omega_n) + g^-({\bf k},\Omega_n)
\nonumber\\
&=& \frac{-2dc{\bf p}^2}{\Omega_n^2 + (c{\bf p})^2}\quad,
\label{eq:4.22}
%\tag{2.58}
\eea
with $g^{\pm}$ from Eq.\ (\ref{eq:3.12a}).
Using the Eqs.\ (\ref{eqs:3.7}) - (\ref{eq:3.9}) in Eq.\ (\ref{eqs:4.21}), 
all of the resulting terms are
proportional to constants times integrals over products of Goldstone
propagators. We find
\bml
\label{eqs:4.23}
\be
\sigma_{(1,2)}(\Omega) = \frac{-i a^2\Gamma_t^2}{4m_{\rm e}^2\Omega}\,
   I(i\Omega_m \rightarrow \Omega + i0)\quad,
\label{eq:4.23a}
\ee
where
\bea
I(i\Omega_m)&=&\frac{1}{V}\sum_{\bf k} k_x^2\ T\sum_{n_2} \Theta(n_2)\,
   g({\bf k},\omega_{n_2})\,
\nonumber\\
&&\hskip -25pt\times\Bigl[
   g({\bf k},\omega_{n_2}+\Omega_m)+g({\bf k},\omega_{n_2} -\Omega_m)\Bigr]
     \quad,
\label{eq:4.23b}
%\tag{2.59a}
\eea
and
\be
a = T\sum_n {\cal E}^{\pm}_{nn}({\bf k}=0)\,N^{(1)}_n/B_n^2\quad.
\label{eq:4.23c}
\ee
\eml%
Notice that ${\cal E}^+_{nn}({\bf k}=0) = {\cal E}^-_{nn}({\bf k}=0)$,
which follows from Eqs.\ (\ref{eqs:3.8}).

Simple asymptotic analysis shows that $I(\Omega) - I(0)\propto
\Omega^{d/2}$. However, before we discuss the integral in detail, the
prefactor, denoted by $a$ in Eq.\ (\ref{eq:4.23a}), warrants a closer
look.

\subsubsection{The prefactor of the nonanalyticity}
\label{subsubsec:IV.B.3}
 
We now discuss the prefactor $a$ defined in Eq.\ (\ref{eq:4.23c}). To
leading order in a disorder expansion, it suffices to calculate $a$
in the clean limit. Simple manipulations yield
\bea
a&=&\frac{T}{2}\sum_n\frac{1}{V}\sum_{\bf k}\left[
    \left({\cal G}_n^+({\bf k})\right)^2 k_x\,\frac{\partial}{\partial k_x}\,
    {\cal G}^-_n({\bf k}) \right.
\nonumber\\
&&\hskip 40pt \left. -\left({\cal G}_n^-({\bf k})\right)^2 k_x\,
   \frac{\partial}{\partial k_x}\, {\cal G}^+_n({\bf k}) \right]\quad.
\label{eq:4.24}
\eea
In the saddle-point approximation for the 
${\cal G}_n^{\pm}$, Eq.\ (\ref{eq:2.13c}), that we have employed so far,
one finds $a=0$. However, this is an artifact of that approximation. To
see this, we first point out that a necessary and sufficient condition
for $a\neq 0$ is a wavenumber dependence of the `magnetic' piece of the
self energy, $\Delta_n$ in Eq.\ (\ref{eq:2.13c}), which is 
${\bf k}$-independent in our saddle-point approximation. Such a 
${\bf k}$-dependence indeed arises from our formalism if we keep the
massive $\delta{\bar\Lambda}$-fluctuations that we neglected so far for
simplicity.\cite{self_energy_footnote} 
Consider Eq.\ (\ref{eq:4.6b}) again. By keeping 
$\delta{\bar\Lambda}$ in the decomposition of $\wt\Lambda$ into
$\bar\Lambda$ and $Q$, Eq.\ (\ref{eq:3.4}), and lumping it into the
Green function, we obtain, upon integrating over $\delta{\bar\Lambda}$
in Gaussian approximation,
a generalization of ${\cal G}^{\pm}$ in Eq.\ (\ref{eq:4.24}) which
has a ${\bf k}$-dependent magnetic self-energy. This procedure still
constitutes an approximation, as determining the prefactor exactly would
require an {\em exact} treatment of the massive modes. It serves to 
demonstrate, however, that the prefactor is in general nonzero, and our
initial null result indeed derives from an oversimplifying approximation.
To avoid misunderstandings, we also point out that the leading frequency
dependence we determine {\em is} exact, and only the prefactor we cannot
calculate exactly.

The above procedure produces a wavenumber dependent self-energy, and hence
a nonzero $a$, even for our model with a point-like spin-triplet interaction
amplitude $\Gamma^{(t)}$, Eq.\ (\ref{eq:2.10c}). A wavenumber dependent
$\Gamma^{(t)}$ would of course also lead to a wavenumber dependence of the
self-energy, but capturing this effect would require a generalization of
our matrix fields. We conclude that the prefactor $a$ in Eq.\ (\ref{eq:4.23a})
is in general nonzero and nonuniversal, as it depends on microscopic details
like the precise strucure of the interaction amplitude.

\subsubsection{The FMGM-induced nonanalyticity}
\label{subsubsec:IV.B.4}

We finally need to consider the integral $I(i\Omega_m)$ defined in
Eq.\ (\ref{eq:4.23b}). At $T=0$, the relevant dimensionless integral
is
\bml
\label{eqs:4.25}
\bea
J(\Omega)&=&\int_0^{\infty} dk\,k^{d+1} \int_0^{\Omega_0} d\omega\ 
   \frac{k^2}{\omega^2 + k^4}\ 
                \left[\frac{k^2}{(\omega - \Omega)^2 + k^4}\right.
\nonumber\\
&&\hskip 60pt \left. + \frac{k^2}{(\omega + \Omega)^2 + k^4}\right]
     \quad.
\label{eq:4.25a}
\eea
First consider the integral in $d=2$. Subtracting the value at $\Omega=0$,
it is easy to see that the leading frequency dependence is linear, i.e.
there is {\em no} term $\propto\Omega\ln\Omega$. More generally,
standard asymptotic analysis yields
\be 
J(\Omega\rightarrow 0) = J(0)
    - \frac{\pi^2}{2^{3+d/2}\sin(\pi d/4)}\ \Omega^{d/2}\quad.
\label{eq:4.25b}
\ee
Consistent with the absence of a logarithm in $d=2$, the prefactor
of the $\Omega^{d/2}$ nonanalyticity is finite in the limit $d\rightarrow 2$. 
\eml%
The frequency scale for the nonanalyticity is given by the quantity
$\Delta$, Eq.\ (\ref{eq:2.15}), which is proportional to the magnetization.

At nonzero temperature, one finds the same qualitative behavior, since 
the Goldstone modes, which are the source of the nonanalyticity, have
the same functional form at all values of $T$, as long as one stays in
the magnetic phase.\cite{finite_t_footnote} 
This is in sharp contrast to the diffuson-induced
weak-localization nonanalyticities, which are cut off by a nonzero
temperature since the diffusons acquire a mass at $T>0$. Collecting
everything, we obtain our final result,
\be
{\rm Re}\,\sigma(\Omega\rightarrow 0) = \sigma_{\rm B}\,\left[1 +
   \frac{C_d}{\epsilon_{\rm F}\tau}\,(\Omega/\Delta)^{(d-2)/2}
     \right]\quad,
\label{eq:4.26}
\ee
which is valid for $2\leq d < 4$. $C_d>0$ in Eq.\ (\ref{eq:4.26}) is a positive
constant (for fixed dimensionality $d$) which is nonuniversal; it depends, 
for instance, on the wavenumber dependence
of the spin-triplet interaction amplitude as was discussed in the
preceding subsection. The $d$-dependence of $C_d$ is nonsingular, in
particular, $C_2$ is a finite number.

\section{Discussion}
\label{sec:V}

In this paper we have developed a general theoretical framework for disordered
itinerant ferromagnets. While microscopic details like band structure etc.
could be built into the theory (they would enter in the form of more
complicated Green functions), our method is particularly well suited for
studying universal properties that are due to the soft modes in the system
and that are independent of the details on microscopic scales. We have
therefore restricted ourselves to one parabolic band, and have studied
the influence of the soft modes on the transport and thermodynamic
properties. Our most important conclusion is that, in disordered systems
with ferromagnetic long-range order, there are two distinct families
of soft modes that contribute to the leading nonanalyticities that lead
to generic scale invariance. One family consists of the
diffusive ``diffusons'' that also exist in the absence of magnetic long-range 
order, while the other are the magnetic Goldstone modes
or spin waves that are characteristic of magnetically ordered systems.
Even though, there are crucial differences between the effects of the
two types of soft modes, which we now discuss in some detail.

As we have seen in Sec.\ \ref{sec:IV}, the one-loop correction to the DOS 
(Fig.\ \ref{fig:4.1}(a)) and the simple loop contribution to the
one-loop correction to the Boltzmann 
conductivity (the first diagram in Fig.\ \ref{fig:4.2}) are simply a
frequency-momentum integral over a soft propagator. For interacting
electrons, the interacting part of the diffuson propagator, i.e. $E$ 
in Eq.\ (\ref{eq:3.8a}) in the diffuson channels, is essentially a 
diffusion propagator squared, i.e., it scales like 
$1/\omega ^{2}\sim 1/{\bf k}^4$. The integration then leads to a term
proportional to $\Omega^{(d-2)/2}$ in $2<d<4$, and $\ln\Omega$ in
$d=2$. This is the usual weak-localization 
contribution.\cite{Cooperon_footnote} 
The FMGM contribution, on the other hand,
consists of a frequency-momentum integral over a {\em single} propagator
that scales like $1/\omega \sim 1/{\bf k}^2$, and is hence less singular.
As a result, there is no FMGM contribution to the leading nonanalyticity
in the DOS.

For the ``football'' contribution to the conductivity (the second diagram
in Fig.\ \ref{fig:4.2}) the situation is different, since both internal 
propagators can be Goldstone modes. Our explicit calculation has shown 
that this term indeed contributes to the leading, $O(\Omega^{(d-2)/2})$, 
nonanalytic frequency dependence of the conductivity. In $d=3$, the
contributions from the diffusons and the FMGMs are thus qualitatively
the same. However, in $d=2$ the diffusons lead to the well-known
``weak-localization'' logarithmic frequency dependence, while the FMGM
contribution does {\em not} lead to a logarithm. A related feature is
that the prefactor of the nonanalyticity in $d=2+\epsilon$ goes like
$1/\epsilon$ for the diffuson-induced term, but is of $O(1)$ in the
FMGM case.

An interesting technical feature is the fact that in the case of the
diffusons, both the simple loop and the ``football'' contribute an
$O(\Omega^{(d-2)/2})$ dependence of the conductivity, while for the
FMGM only the latter does. As mentioned above, the different results
from the simple loop can be understood by realizing that the interacting
part of the diffuson is really a diffusion propagator
squared that scales like $1/{\bf k}^4$, while the FMGM scales like
$1/{\bf k}^2$. A question that arises then is why the diffusons in the
``football'' do not yield a much stronger singularity, of 
$O(\Omega^{(d-6)/2})$. The technical reason is that the diffusons
come with additional frequency restrictions, as was discussed in
Sec.\ \ref{subsec:III.B}. These frequency restrictions lead to
additional frequency factors in the integral that determines the
``football'' contribution, while no such singularity-protecting effect
occurs in either the simple loop for the diffusons or in any of the
FMGM contributions.

Another very important difference between diffusons and FMGMs is that
the former are soft only at $T=0$, while the latter remain soft at all
$T$ below the Curie temperature. As a result, the nonanalytic frequency
dependence coming from the diffusons gets replaced by an analogous
nonanalytic temperature dependence at $T>0$. The one induced by the
FMGM, on the other hand, is still of $O(\Omega^{(d-2)/2})$ even at
$T>0$.\cite{finite_t_footnote}

The sign and strength of the nonanalytic term in Eq.\ (\ref{eq:4.26}) 
is interesting and can be understood on very general grounds. First 
we remember that this
nonanalyticity in frequency also exists at finite temperature. This implies
that its origin is classical in nature. The sign of classical ``mode
coupling'' or ``generic scale invariance'' contributions to the transport
coefficients is determined by the structure of the hydrodynamic equations
describing the long wavelength dynamics of the soft modes.\cite{Kawasaki}
For the spin
density dynamics relevant here, the leading nonlinear coupling in the long
wavelength limit contains two gradients, as it does in classical
dissipative systems such as Lorentz models.\cite{Hauge} 
As in the Lorentz models,
the soft modes therefore have a localizing effect, i.e. the conductivity
decreases with decreasing frequency. In contrast, in classical fluids
the leading nonlinear coupling contains only one gradient, and the
prefactor of the nonanalyticity has a delocalizing sign.\cite{LTT}
We further note that the nonanalyticity in Eq.\ (\ref{eq:4.26}) is stronger
than the one in classical Lorentz gases by a factor of $\Omega$. It is,
in fact, as strong as the generic
scale invariance effects in classical fluids, even though the structures
of the respective nonlinear
couplings suggest just the opposite. The resolution of this apparent
paradox lies in the fact that there is long-range order in our system,
which is manifested by the Goldstone modes. In a mode-coupling calculation,
the transverse spin suspectibilty, $\chi_t \propto 1/{\bf k}^2$, appears as 
a multiplicative factor in the integrands and effectively leads to the same 
power of the frequency as in the fluid case. It is interesting to note that 
a similar mechanism in certain liquid crystals
leads to mode coupling effects that are even stronger than in classical
fluids.\cite{MRT} For these systems the standard hydrodynamic description
breaks down for all $d<5$. Here, the analogous critical dimension is
$d=2$, see Eq.\ (\ref{eq:4.26}).

There are some important experimental consequences of these results.
One is that, as far as the weak-localization properties in $d=2$
are concerned, one does not expect any differences between metals with
long-range ferromagnetic order and nonmagnetic metals in an external
magnetic field. The reason is that the only difference in the soft
mode structures of the two systems are the Goldstone modes,
but they do not contribute to the leading nonanalyticities
in $d=2$. Indeed, transport measurements on thin ferromagnetic films have
found just the ordinary weak-localization effects due to the diffusons,
with no additional effects from the Goldstone modes at all.\cite{Giordano}
Our theory explains this null result, which {\em a priori} is very surprising.
For bulk materials, on the other hand, our theory predicts that the FMGMs
do contribute a term that is qualitatively the same as the weak-localization
effect, and furthermore that this term will remain a nonanalytic frequency
dependence even at $T>0$.

Another consequence is the suggestion that the FMGMs will
not be important for the critical behavior at the metal-insulator transition
from a FM metal to a FM insulator, at least close to two-dimensions. The
reason for this is the absence of a diverging prefactor of the frequency
nonanalyticity as $d\rightarrow 2$. This divergence is known to drive the
transition in $d=2+\epsilon$, and since the FMGMs do not contribute to it
one expects the universality class of the transition to be unchanged by
the presence of ferromagnetic long-range order.\cite{high_d_footnote}
We will investigate this point in (II)
and find that the metal-insulator transition on the background of
ferromagnetism is indeed closely related to the one of nonmagnetic electrons
in an external magnetic field.

\acknowledgments
We gratefully acknowledge stimulating discussions with Nick Giordano.
Parts of this work were performed at the Institute for Theoretical Physics
at UC Santa Barbara, and at the Aspen Center for Physics. This work was 
supported by the NSF under grant Nos. DMR-98-70597, DMR-99-75259, 
and PHY94-07194.


\begin{references}
\b{BaymPethick} See, e.g., G. Baym and C. Pethick, {\it Landau Fermi--Liquid
 Theory}, John Wiley (New York 1991).
\b{FetterWalecka} A.L. Fetter and J.D. Walecka, {\it Quantum Theory
 of Many--Particle Systems}, Mc Graw--Hill (New York 1971).
\b{AGD} A.A. Abrikosov, L.P. Gorkov, and I.E. Dzyaloshinskii, {\it Methods
 of Quantum Field Theory in Statistical Physics} (Dover, New York, 1975).
\b{weak_localization_footnote} We use the term ``weak localization'' to refer
 to the nonanalytic behavior of electronic correlation functions in the limit
 of zero momentum and/or frequency that is induced by diffusive soft modes
 in interacting disordered electron systems. In the literature, the term is
 often restricted to those effects that survive in the limit of noninteracting
 electrons. 
\b{LeeRama} For reviews, see, e.g., P.A. Lee and T.V. Ramakrishnan, 
 Rev. Mod. Phys. {\bf 57}, 287
 (1985); B.L. Altshuler and A.G. Aronov, in {\it Electron-Electron Interactions
 in Disordered Systems}, M. Pollak and A.L. Efros (eds.) (North Holland,
 Amsterdam, 1984), p.1.
\b{Shankar} R. Shankar, Rev. Mod. Phys. {\bf 66}, 129 (1994).
\b{Shankar_ff} Following Ref.\ \onlinecite{Shankar}, a large
 number of follow-ups, generalizations, and alternatives have been put
 forward, see Ref.\ \onlinecite{us_fermions} and references therein.
\b{us_fermions} D. Belitz and T.R. Kirkpatrick, Phys. Rev. B {\bf 56}, 6513
 (1997); D. Belitz, F. Evers, and T.R. Kirkpatrick, Phys. Rev. B {\bf 58},
 9710 (1998).
\b{GSI} See, e.g., B.~M. Law and J.~C. Nieuwoudt, Phys. Rev. A
 {\bf 40}, 3880 (1989); J.~R. Dorfman, T.~R. Kirkpatrick, and J.~V. Sengers,
 Ann. Rev. Phys. Chem. {\bf 45}, 213 (1994); S. Nagel, Rev. Mod. Phys.
 {\bf 64}, 321 (1992). The possibility of interpreting weak-localization 
 effects as generic scale invariance phenomena has been emphasized by
 T.R. Kirkpatrick and D. Belitz, J. Stat. Phys. {\bf 87}, 1307 (1997).
\b{us_paper_II} T.R. Kirkpatrick and D. Belitz, to be published.
\b{NegeleOrland} J. W. Negele and H. Orland, {\it Quantum Many-Particle
 Systems} (Addison-Wesley, New York, 1988).
\b{Notation_Footnote} We use the notation $a\cong b$ for "$a$ is isomorphic to
 $b$", $a\propto b$ for ``$a$ is proportional
 to $b$'', and $a\sim b$ for ``$a$ scales like $b$''.
\b{dis_footnote} A second contribution to ${\cal A}_{\rm dis}$, which is
 given in terms of $(\tr Q)^2$ rather than $\tr (Q^2)$,\cite{us_fermions}
 is irrelevant for the purposes of this paper.
\b{Moriya} T. Moriya, {\it Spin Fluctuations in Itinerant Electron Magnetism}
 (Springer, Berlin, 1985).
\b{R} For a review, see, e.g., D. Belitz and T.R. Kirkpatrick, Rev. Mod. Phys.
 {\bf 66}, 261 (1994).
\b{IKK} T. Izuyama, D.J. Kim, and R. Kubo, J. Phys. Soc. Jpn. {\bf 18}, 1025
 (1963).
\b{Kubo} R. Kubo, J. Phys. Soc. Japan {\bf 12}, 570 (1957).
\b{Mahan} G.D. Mahan, {\it Many-Particle Physics}, Plenum (New York, 1986),
 Sec. 3.7.
\b{CastellaniKotliar} C. Castellani and G. Kotliar, Phys. Rev. B {\bf 36},
 7407 (1987).
\b{cds_trk} C.V. De Souza and T.R. Kirkpatrick, Phys. Rev. B {\bf 43}, 10132
 (1991).
\b{self_energy_footnote} Alternatively, it is easy to see by means of standard
 diagrammatic many-body theory that the exact self-energy for our model is
 wavenumber dependent.
\b{finite_t_footnote} This holds for the processes considered here. There
 may be additional processes at $T>0$ due to the fact that thermally excited
 spin waves exist. This will be the subject of a separate investigation.
\b{Cooperon_footnote} In nonmagnetic systems, in the absence of an external
 magnetic field and magnetic impurities, there is another
 mechanism for producing such terms. In
 such systems, the particle-particle or Cooper channel ($r=1,2$) is also
 diffusive, and perturbation theory yields a simple momentum integral over
 a diffusive Cooperon, leading again to a term proportional to
 $\Omega^{(d-2)/2}$.
\b{Kawasaki} K. Kawasaki, J. Phys. A {\bf 6}, L1 (1973).
\b{Hauge} See, e.g., E.H. Hauge, in {\it Transport Phenomena}, Lecure Notes
 in Physics No. 31, edited by G. Kirczenow and J. Marro, Springer (New
 York 1974), p.337.
\b{LTT} See, e.g., J.R. Dorfman et al, Ref.\ \onlinecite{GSI}.
\b{MRT} G.F. Mazenko, S. Ramaswamy, and J. Toner, Phys. Rev. A {\bf 28}, 1618
 (1983).
\b{Giordano} N. Giordano, private communication.
\b{high_d_footnote} We note, however, that this is true only in theoretical
 descriptions of the metal-insulator transition near $d=2$. In high dimensions
 ($d>6$), the nature of the transition is not related to the (very weak)
 nonanalyticities in the metallic phase, see T.R. Kirkpatrick and D. Belitz,
 Phys. Rev. Lett. {\bf 73}, 862 (1994).
\end{references}
\end{document}